\documentclass[a4paper,fleqn]{cas-dc}

\usepackage[numbers]{natbib}

\def\tsc#1{\csdef{#1}{\textsc{\lowercase{#1}}\xspace}}
\tsc{WGM}
\tsc{QE}
\tsc{EP}
\tsc{PMS}
\tsc{BEC}
\tsc{DE}

\usepackage{graphicx} 
\usepackage{dcolumn} 
\usepackage{bm} 
\usepackage{amsmath}
\usepackage{amsopn}
\usepackage{amsfonts}
\usepackage{amssymb}
\usepackage{mathtools,nccmath}
\usepackage{float}
\usepackage{epstopdf} 
\usepackage{xcolor}
\usepackage{natbib}
\usepackage{stfloats}

\usepackage{caption}
\usepackage{multicol}

\newcommand{\tblmultiline}[2][c]{\begin{tabular}[#1]{@{}c@{}}#2\end{tabular}}

\begin{document}
\let\WriteBookmarks\relax
\def\floatpagepagefraction{1}
\def\textpagefraction{.001}
\shorttitle{Ultra-high lattice thermal conductivity...}
\shortauthors{Nayeb Sadeghi... et~al.}

\title [mode = title]{Ultra-high lattice thermal conductivity and the effect of pressure in superhard hexagonal BC$_{2}$N}                      



\author[1]{Safoura Nayeb Sadeghi}[type=editor,
                        auid=000,bioid=1,
                         prefix=,
                          role=,
                        orcid=0000-0000-0000-0000]
\ead{safoura@virginia.edu}


\address[1]{Department of Physics, University of Tehran, Tehran, Iran}

\author[1]{S.Mehdi Vaez Allaei}[%
   ]
 \cormark[1]
\ead{smvaez@ut.ac.ir}

\author[2,3]{Mona Zebarjadi}[%
   ]
\ead{mona.zebarjadi@virginia.edu}


\address[2]{Department of Electrical and Computer Engineering, University of Virginia, Charlottesville, VA, USA.}
\address[3]{Department of Materials Science and Engineering, University of Virginia, Charlottesville, VA, USA.}

\author%
[4,3,5]
{Keivan Esfarjani}
\cormark[2]
\ead{k1@virginia.edu}

\address[4]{Department of Mechanical Engineering, University of Virginia, Charlottesville, VA, USA.}
\address[5]{Department of Physics,University of Virginia, Charlottesville, VA, USA}

\cortext[cor1]{Corresponding author}
\cortext[cor2]{Corresponding author}



\begin{abstract}
Hexagonal BC$_{2}$N is a superhard material recently identified to be comparable to or even harder than cubic boron nitride (c-BN) due the full $sp^3$ bonding character and the higher number of C-C and B-N bonds compared to C-N and B-C. 
Using a first-principles approach to calculate force constants and an exact numerical solution to the phonon Boltzmann equation, we show that BC$_{2}$N has a high lattice thermal conductivity exceeding that of c-BN owing to the strong C-C and B-N bonds, which produce high phonon frequencies as well as high acoustic  velocities. The existence of large group velocities in the optical branches is responsible for its large thermal conductivity.
Its  coefficient of thermal expansion (CTE) is found to vary from 2.6$\times$10$^{-6}$/K at room temperature to almost 5$\times$ 10$^{-6}$/K at 1000K. The combination of large thermal conductivity and a good CTE match with that of Si,  makes BC$_2$N  a promising material for use in thermal management and high-power electronics applications. 
We show that the application of compressive strain increases the thermal conductivity significantly. This enhancement results from the overall increased frequency scale with pressure, which makes acoustic and optic velocities higher, and weaker phonon-phonon scattering rates.


\end{abstract}

\begin{keywords}
Hexagonal BC$_2$N  \sep Thermal conductivity \sep First principles \sep 
\end{keywords}

\maketitle

\section{Introduction}

Supermaterials can deliver extreme performance across a variety of applications. In particular, diamond, a covalent insulator with simple electronic and lattice structures, has long been known as one of the hardest ~\cite{teter1998computational} and highest thermally conductive materials \cite{onn1992some,olson1993thermal,wei1993thermal} on earth due to its strong $sp^3$ covalent bonds and the light mass of carbon atoms. However, its inherent shortcomings such as brittleness, oxidization, and reaction with iron, and difficuly in synthesis, restrict its practical applications to some degree\cite{chung2007synthesis,kaner2005rb}. As a consequence, the pursuit of superhard thermally conductive materials with desirable properties is an recurring theme in condensed matter physics and materials science. Boron and Nitrogen are the adjacent elements to Carbon and have a similar atomic radius. Their binary phases or elemental solids containing B or N dopant exhibit excellent mechanical performance and high stability. For instance, cubic BN with an acceptable hardness value shows better chemical and thermal stability (approximately 1650 K) and ductility in comparison with diamond \cite{wentorf1957cubic}. In terms of its thermal performance, experimental and theoretical studies have been conducted to investigate thermal transport in both naturally occurring and isotopically pure c-BN, revealing that it possesses one of the highest thermal conductivities reported but lower than that of diamond. Recently, Lindsay et al.\cite{lindsay2013first}, and S.Mukhopadhyay et al. \cite{mukhopadhyay2014polar} estimated similar room temperature values of (2145 and 2113 W/m-K for $\kappa_{RT}$ from first-principles calculations based on a real space approach and DFPT, respectively. These were greater than the experimental value of 1300 W/m-K \cite{slack1973nonmetallic} (due to the absence of scatterings from boundaries and impurities), and lower than the reported value for diamond ($>$ 3000 W/m-K) using a similar approach \cite{ward2009ab}. However, the hardness of c-BN is only in the range of 46-66 GPa\cite{zhao2002superhard,solozhenko2001synthesis,luo2007body,solozhenko2001mechanical}, much lower than that of diamond which is 100-160 GPa \cite{solozhenko2001mechanical}. More recently, Chakraborty et al. \cite{chakraborty2018lattice} revealed that Hexagonal diamond (h-C) and wurtzite boron nitride (w-BN) are two superhard materials comparable to or even harder than their cubic counterparts, cubic diamond (c-C) and cubic boron nitride (c-BN). By using first-principles calculations, they also estimated their lattice thermal conductivity ($\kappa_{L}$) exceeding the overall thermal conductivity of metals, albeit lower than that of their cubic counterparts due to their larger volume of three-phonon scattering phase space. Therefore, the search for new superhard thermally conductive materials with properties comparable or even superior to those of diamond is still an intriguing and long-standing endeavor. Similarities between diamond and cubic boron nitride (c-BN), such as their closely matched lattices, high melting temperatures ($>$3000 K), large bulk moduli, high thermal conductivity, and similar thermal-expansion coefficients, motivate the expectation that zinc-blende-structured cubic boron-carbonitride (c-BCN) compounds may form new superhard and superabrasive materials. Therefore, considerable efforts have been devoted to synthesizing various B-C-N compounds with covalent and rigid network lattices under high pressure and temperature. Up to now, several  successful syntheses of ternary B-C-N compounds (e.g., BCN \cite{luo2006synthesis,huang2004synthesis,kaner1987boron,yamada1998shock}, c-BC$_{x}$N (x=0.9-3.0) \cite{knittle1995high}, BC$_{2}$N \cite{nakano1994segregative,komatsu1996synthesis,solozhenko2001synthesis,zhao2002superhard,yao1998amorphous,tkachev2003elastic,solozhenko2001mechanical,wu2006thermal,kouvetakis1989novel}, BC$_{4}$N \cite{solozhenko1997phase,hubavcek1995preparation,zhao2002superhard}, and BC$_{6}$N, B$_{2}$CN \cite{he2001orthorhombic}) have been achieved by using different methods such as chemical vapor deposition \cite{kaner1987boron}, solvothermal method\cite{huang2004synthesis}, chemical process method\cite{hubavcek1995preparation}, high temperature and high pressure (HTHP) method\cite{zhao2002superhard,solozhenko2001synthesis,solozhenko2001mechanical,nakano1994segregative}, mechanical milling method\cite{yao1998amorphous}, spark plasma sintering\cite{luo2006synthesis}, and shock-compression
method\cite{komatsu1996synthesis}. Among them, BC$_{2}$N has drawn extensive attention since it has been expected to be thermally and chemically more stable than diamond and harder than c-BN. The measured Vickers hardness of the recently synthesized c-BC$_{2}$N reached 76 or 62 GPa\cite{solozhenko2001synthesis,solozhenko2001mechanical,zhao2002superhard}indeed higher than c-BN. However, the determination of its structure becomes rather difficult owing to the small and similar atomic masses of B, C, and N. This inspires theoretical researchers to resolve this challenging problem\cite{kim2007cubic,mattesini2001search,zhang1999energetics,pan2005ab,mattesini2001first,widany1998density,zhang1999energetics,lambrecht1989electronic,tateyama1997proposed}. For instance, on the theoretical side, many different crystal forms were proposed, e.g. zinc-blende structures \cite{sun2001structural,sun2006first,zhang2004superhard}, chalcopyrite (cp-) BC$_{2}$N \cite{sun2006chalcopyrite}, body-centered (bc6-) BC$_{2}$N \cite{luo2007body}, tetragonal z-BC$_{2}$N \cite{zhou2007most}, wurtzite BC$_{2}$N \cite{luo2007first}, the short period (C$_{2}$)$_{n}$(BN)$_{n}$ (111) superlattices \cite{chen2007superhard,chen2008crystal,chen2007chen}, and tetragonal (t-) BC2N \cite{zhou2009tetragonal}. Among the proposed structures, z-BC$_{2}$N \cite{zhou2007most}  is constructed from the sixteen-atom supercell of diamond and its XRD spectrum agrees with the experiment data \cite{solozhenko2001synthesis} very well, and generally superlattices \cite{chen2007superhard,chen2008crystal}  have among the lowest energies. The t-BC2N phase was predicted to be another candidate because the estimated Vickers hardness and the simulated XRD and Raman patterns were in excellent agreement with the experimental results\cite{zhou2009tetragonal}.  

Recently, Q.Li et al. \cite{li2009rhombohedral} proposed a rhombohedral structure with R3m-2u symmetry as the best candidate for the superhard BC$_{2}$N by the ab initio evolutionary algorithm using USPEX code \cite{oganov2006crystal,glass2006uspex,oganov2006high}, which well reproduced the experimental x-ray diffraction (XRD)  and energy-loss near-edge spectroscopy (ELNES) patterns. They revealed that this structure has lower energy in comparison with earlier-proposed structures e.g struc-1, BC$_{2}$N-w3, z-BC$_{2}$N, and BC$_{2}$N $1\times 1$ and BC$_{2}$N $2\times 2$. They also demonstrated that earlier-proposed high-density and low-density forms are likely from this single rhombohedral phase. The theoretical hardness of R3m-2u BC2N is 62.1 GPa, in excellent agreement with the experimental value of 62 or 76 GPa, which exceeds that of c-BN (58.6 GPa) measured by the same method.

\begin{figure}[pos=H]
\centering
  \includegraphics[height=5 cm]{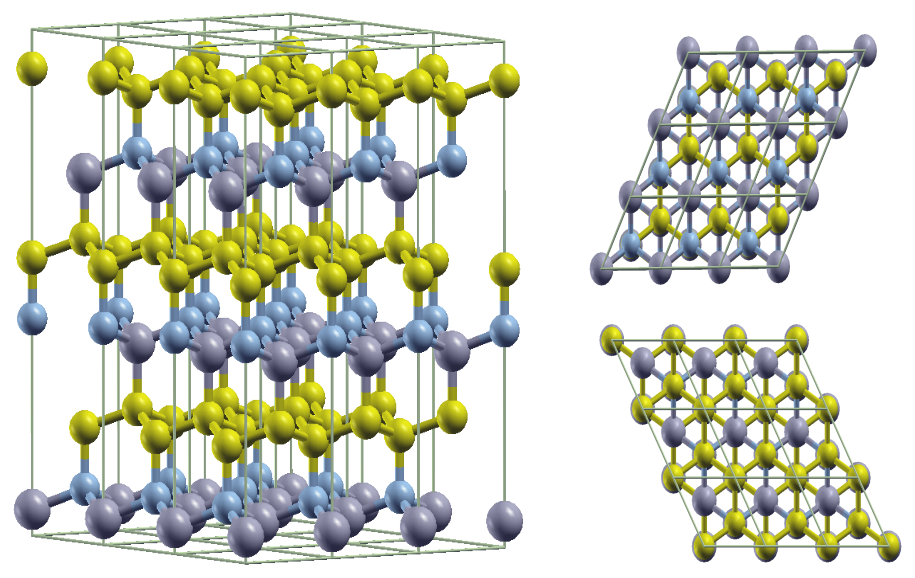}
    \caption{Crystal structure and top views of hexagonal BC$_{2}$N , of space group R3m, proposed by Liu et al. \cite{liu2018hexagonal}. The purple, blue, and yellow spheres represent B, N, and C atoms, respectively.}
  \label{structure}
\end{figure}

More Recently, L.Liu et al.\cite{liu2018hexagonal} proposed R3m BC$_{2}$N with the second highest hardness value of 71 GPa through first-principles swarm structure calculations. The simulated X-ray diffraction patterns, K-edge spectra, and hardness value of R3m BC$_{2}$N well matched with the experimental measurements. Moreover, they revealed that this structure is more stable than the earlier proposed BC$_{2}$N phases, e.g., P4cc, Pmm2, P$\bar{4}$21m, and P$\bar{4}$2m phases. This proposed structure is displayed in Fig. \ref{structure} and the lattice is characterized in table \ref{lattice-parameters}.
It consists of a superposition of  buckled layers of honeycomb C and hexagonal B-N. As it can be seen, the two kinds of honeycomb layers are interconnected through C-N and C-B bonds.

\begin{table}[pos=H]
\caption{Lattice parameters of hexagonal BC$_{2}$N.}
\begin{tabular}{p{2cm} p{.5cm} p{1.2cm} p{.7cm} p{.7cm} p{.7cm}}
\hline\hline                      
\tblmultiline{Lattice \\ parameters} & \tblmultiline{atom \\ type} & \tblmultiline{Wyckoff \\ positions} & \centering x & \hspace{0.1 cm} y &  z \\ [0.5ex]
\hline
\tblmultiline{space group: \\
R3m (No.160)} & B  & 3a & 0.0000 & 0.0000 & 0.0515  \\

\vspace{0.2 cm}
a=b=2.541 \AA & \vspace{0.2 cm} C  & \vspace{0.2 cm}3a  & \vspace{0.2 cm}0.0000 & \vspace{0.2 cm}0.0000 & \vspace{0.2 cm}0.5491  \\
\vspace{0.2 cm}
 c=12.618 \AA & \vspace{0.2 cm} C  & \vspace{0.2 cm}3a & \vspace{0.2 cm} 0.0000 & \vspace{0.2 cm} 0.0000 & \vspace{0.2 cm}0.9221  \\
\vspace{0.2 cm}
  & \vspace{0.2 cm}N  & \vspace{0.2 cm} 3a & \vspace{0.2 cm}0.0000 & \vspace{0.2 cm}0.0000 & \vspace{0.2 cm}0.4280  \\

\hline
\end{tabular}
\label{lattice-parameters}
\end{table}

Moreover, the B or N atoms in the B-N layer project at the center of the carbon hexagonal rings  Fig.\ref{structure}. This structural character effectively avoids the formation of B-B and N-N bonds. The presence of a pure C honeycomb layer, containing strong C-C bonds, obviously indicates that the predicted BC$_{2}$N is the high-density phase. Its unusual hardness can be attributed to the full $sp^3$ bonding character and the higher number of C-C and B-N bonds compared to C-N and B-C. Liu et al. also showed that R3m BC$_{2}$N has a wide indirect electronic band gap around 5.0 eV by using the Heyd Scuseria Ernzerhof (HSE) screened hybrid functional\cite{liu2018hexagonal}. This is slightly smaller than the band gap value (5.47 eV) of diamond. Nonetheless, the thermal properties of BC$_{2}$N were not calculated, inspiring us to investigate its thermal conductivity $\kappa$.  Ab initio calculations based on Boltzmann transport theory in combination with density functional theory (DFT) have demonstrated accuracy in describing thermal conductivity of a variety of bulk materials \cite{broido2007intrinsic,ward2009ab,omini1996beyond,garg2011role,esfarjani2011heat,tang2010lattice} and 2D structures \cite{huang2019thermal,pandey2017ab,zeraati2016highly}. 

So, in this study, we calculated the force constants using ab initio methods, and solved the Boltzmann equation to investigate the thermal conductivity of the most energetically stable structure (R3m) BC$_{2}$N reported so far and compare it to diamond and c-BN.

Moreover, as ultrahard materials are usually used under high-pressure conditions, we also systematically studied how  uniaxial, biaxial and hydrostatic pressure affect thermal transport in this material.

\section{Computational Methods}

The ab initio determination of thermal conductivity and phonon scattering rates are enabled by obtaining geometry and interatomic force constants (IFCs) from density functional theory (DFT) calculations and solving the linearized Boltzmann transport equation (BTE). BC$_{2}$N with R3m space group has been found to be the most energetically stable structure reported so far.  In this work, we perform first  principles  calculations  by using the Quantum Espresso \cite{giannozzi2009quantum} code with the generalized gradient approximation (GGA) and the PBEsol exchange-correlation functional and by using ultrasoft pseudopotentials.  All atoms were fully relaxed with a kinetic-energy  cutoff of  80 Ry and  a  $20\times20\times 4$  Monkhorst-Pack k-mesh until the Hellmann-Feynman forces on them were less than $10^{-6}$ Ry/Bohr.  The applied compressive pressure was simulated by decreasing the lattice constant and relaxing the atoms in the smaller volume structure. The harmonic  and  anharmonic  interatomic  force  constants (IFCs)  were  obtained  based  on  the  direct  supercell  approach. The harmonic IFCs were calculated using a $4\times4\times1$ supercell with a $5\times5\times4$ $\Gamma$-centered Monkhorst-Pack k mesh by using the Phonopy package \cite{phonopy}. Anharmonic IFCs were calculated using 50 Ryd kinetic-energy  cutoffs and gamma point calculations with $4\times4\times1$ supercells, and the cutoff  for cubic IFCs was considered up to the seventh nearest neighbors. After obtaining the  second-order  harmonic  and  third-order  anharmonic IFCs,  the thermal conductivity of naturally  occurring isotopic  and  hypothetical isotopically  pure  hexagonal BC$_{2}$N  were calculated on a converged q-space grid $18\times18\times3$ by iteratively  solving  the  phonon  Boltzmann transport equation as implemented in the Sheng-BTE software  package\cite{li2014shengbte}. Within the framework of the BTE\cite{ziman2010electrons}, and the relaxation time approximation (RTA) the phononic thermal conductivity tensor ($\kappa$) can be expressed as:

\begin{equation}\label{Klattice}
 \kappa^{\alpha\beta}_{\ell} = \frac{1}{V N}\sum_{\lambda}(\hbar\omega_{\lambda})(\frac{\partial f_{\lambda}}{\partial T})\,\upsilon_{\lambda}^{\alpha}\upsilon_{\lambda}^{\beta}\tau_{\lambda}
\end{equation}

where $\alpha$ and $\beta$ are the Cartesian directions, $\lambda$ comprises both a phonon branch index $p$ and a wave vector $q, T$ is the absolute temperature, $V$ is the volume of the unit cell, $N$ is the number of $q$ points in the Brillouin zone, $\hbar$ is the reduced Planck constant, $\omega_{\lambda}$ and $\upsilon_{\lambda}$ are the angular frequency and group velocity of phonon mode $\lambda$ respectively, $f_{\lambda}$ is the equilibrium Bose-Einstein distribution function. The inverse of the phonon lifetime $\tau_{\lambda}$ in the relaxation time approximation (RTA) is equal to the total scattering rate, which is a sum of the anharmonic scattering rates and the isotope scattering rate. The scattering matrix elements $V^{\pm}_{\lambda \lambda' \ \!\!\! \lambda'' \ }$ appearing in the relaxation rates are essentially the Fourier coefficients of the third-order interatomic force constants, rotated to the phonon eigenmode basis. The so called “weighted phase space” $\textbf W$ is defined as the sum over terms involving frequencies in the expression of three-phonon scattering rates Eq.(\ref{em-abs}), which are written as: 

\begin{equation} 
  \textbf{w}^{\pm}_{\lambda} = \frac{1}{2N}\sum_{\lambda' p"}\{^{2(f_{\lambda'}-f_{\lambda"})}_{f_{\lambda'}-f_{\lambda"}+1}\}\frac{\delta(\omega_{\lambda}\pm\omega_{\lambda'}-\omega_{\lambda"})}{\omega_{\lambda} \omega_{\lambda' \ } \omega_{\lambda'' \ }}
  \label{em-abs}
\end{equation}
for absorption (+) and emission (-) processes\cite{li2014thermal,li2015ultralow}. It does not involve the cubic force constants, but shows the volume in the phase space where energy and momentum are conserved in a three-phonon process. The larger the scattering phase space, the larger the three-phonon scattering rate.  We also calculated the elastic constants and thermal expansion by using thermo-pw\cite{palumbo2017lattice} code implemented as a utility of QE to compute thermodynamic properties. 

\begin{figure}[pos=H]
    \centering
    \includegraphics[height=5 cm]{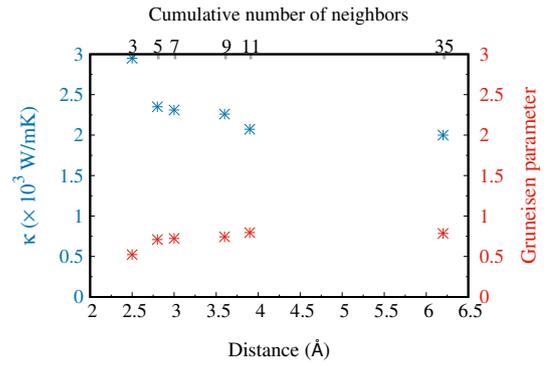}
    \caption{Thermal conductivity and Gruneisen parameters of the unstrained sample at room temperature versus the number of neighbors included in cubic anharmonic force constants. Results seem to converge after 11 neighbors.}
    \label{convergence}
\end{figure}

\underline{\bf{Convergence studies}}:

To study the convergence with respect to the cutoff range of the cubic force constants, we plot in Fig. \ref{convergence} the final thermal conductivity and average Gruneisen parameter at room temperature and under zero pressure ,as a function of the cutoff distance.

It can be inferred that the results converge for 11 neighbors, corresponding to a cutoff distance of 4 \AA.
In our calculation of thermal properties we have however used and reported the results with 7 neighbors. One can see that the converged values of the Gruneisen parameters are approximately 7\% larger and the converged thermal conductivities are 15\% lower. This does not change the results of our discussions and figures in the paper as we are mostly discussing trends and relative contributions. The only exception being the absolute thermal conductivities in Fig. \ref{thermal-conductivity} and scattering rates in Fig. \ref{anharmonic-SRs} where they are plotted on a log scale. A 15\% change is however invisible on this scale. So all figures present in this paper remain to a great degree unchanged if more neighbors were to be included. 

\begin{figure}[pos=H]
\centering
  \includegraphics[height=6.3 cm]{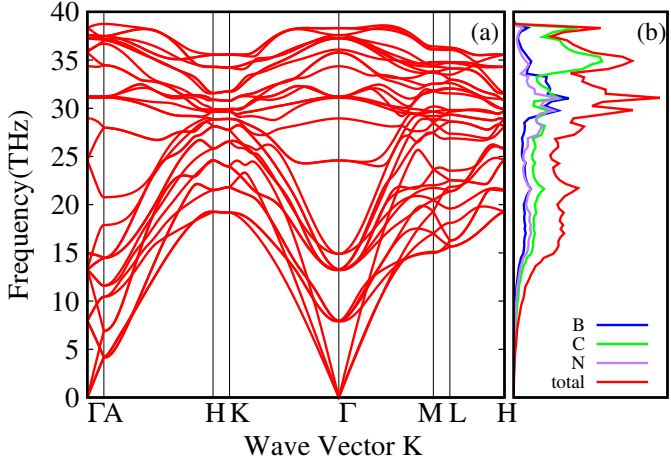}
    \caption{Phonon dispersion curves (a), and phonon density of states (b) with contributions from each element of hexagonal BC$_{2}$N.}
  \label{phonon-dispersion}
\end{figure}

\section{Results}
\subsection{Phonon dispersion and elastic properties of hexagonal BC$_{2}$N}

The phonon dispersion in the equilibrium geometry, Fig.\ref{phonon-dispersion}(a), is in good agreement with the first-principles calculations reported in Ref.\cite{liu2018hexagonal}. The corresponding phonon density of states with contributions from each element of hexagonal BC$_{2}$N is shown in Fig.\ref{phonon-dispersion}(b). Phonon modes at frequencies higher than 32 THz are dominantly due to vibrations of C atoms (stiffer C-C bonds), while below this frequency the three elements have almost equal contributions.
As can be seen in Fig.\ref{phonon-dispersion}(a), the frequencies of all phonon modes are positive, which reveals the dynamical stability of hexagonal BC$_{2}$N. The elastic constants of hexagonal BC$_{2}$N at 0 GPa were also calculated. 
For a hexagonal structure, there are five independent elastic constants. The calculated elastic constants are given in table \ref{elasticonstants} in comparison with previous theoretical studies for hexagonal BC$_{2}$N, c$-$BN, and diamond. 
On the basis of Born stability criteria, the elastic constants for a hexagonal structure should satisfy the following elastic stability criteria: C$_{44}>0$, C$_{11}-$ C$_{12} >0$, and C$_{33}$(C$_{11}$+C$_{12}$)-2C$^2_{13} > 0$ ( $C_{66} = (C_{11}-C_{12})/2$ ) \cite{mouhat2014necessary}. 
These results reveal that the Hexagonal BC$_{2}$N is mechanically stable. In the Voigt–Reuss–Hill (VRH) approximation which models elastic properties of polycrystalline materials, the bulk modulus (B), shear modulus (G), Young modulus (E), and Poisson ratio ($\nu$) are also given in table \ref{elasticonstants}. 
According to Pugh ’s rule, ductile behavior is exhibited in materials when B/G$>$1.75, otherwise the materials behave in a brittle manner. For BC$_{2}$N, B/G=0.90$<$1.75, which shows it may be brittle. 
According to the rule proposed by Frantsevich, a brittle behavior is exhibited in materials when $\nu<$0.26, otherwise the materials behave in a ductile manner. 
For BC$_{2}$N, $\nu$=0.094$<$0.26, which is in agreement with Pugh's rule. 
All elastic constants and elastic moduli values are consistent with previous theoretical studies \cite{liu2018hexagonal,li2019elastic}. 
The results of these studies for the Vickers hardness are 70.34 GPa, and 71 Gpa, which are close to the experimental value of 76 GPa. They have also shown that the hardness increases notably with compressive strain \cite{li2019elastic}. The obtained moduli and hardness values are intermediate between that of c-BN and diamond \cite{mounet2005first}.

\begin{table*}
\centering

\caption{Equilibrium lattice constants (a and c, in \AA ), space group (sg), elastic constants (C$_{ij}$, in GPa), elastic moduli (B, G, E, in GPa), and Poisson's ratio of hexagonal BC$_{2}$N, cubic BN, and diamond.}

\begin{tabular}{|p{1.6cm}| p{.9cm}| p{.8cm} | p{.8cm}| p{.9cm}| p{.9cm}| p{.7cm}| p{.9cm}| p{.9cm}| p{.9cm}| p{.7cm}| p{.7cm}| p{.8cm}| p{.7cm}|}

\hline\hline                      
   & $sg$ & $a$ &$c$&C$_{11}$&C$_{12}$&C$_{13}$&C$_{33}$&C$_{44}$&C$_{66}$&$B$&$G$&$E$&$\nu$ \\[0.5ex]

\hline


Present & R3m & 2.528 & 12.557 & 1049 & 112.32 & 65.18 & 1054.52 & 406.97 & 468.39 & 404 & 450 & 985 & 0.094  \\
\hline
BC$_{2}$N\cite{liu2018hexagonal} & R3m & 2.541 & 12.618 & 1024.76 & 103.57 & 61.17 & 1022.42 & 402.55 \\ 
\hline
BC$_{2}$N\cite{li2019elastic} & R3m & 2.538 & 12.603 & 1025 & 99 & 56 & 1024 & 405 & 463 & 338 & 444 & 965 & 0.08 \\
\hline
c$-$BN\cite{cao2019superhard} & F-43m& 3.622& & 778 & 165 & & & 447& &369 &384 &856& 0.11 \\
\hline
diamond\cite{fan2015structural} &F-43m& 3.620& & 1053& 120& & & 563& & 432& 522& 1116& 0.07\\[1ex]
\hline
\end{tabular}

\label{elasticonstants}
\end{table*}

\begin{figure}[pos=H]
 \centering
 \includegraphics[height=7 cm, angle=-90]{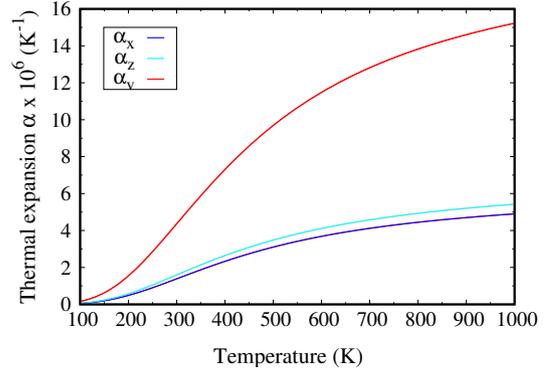}
  \caption{Coefficients of volume and linear thermal expansion as a function of temperature for hexagonal BC$_{2}$N within QHA. Note they are related through $\alpha_V=\alpha_z+2\alpha_x$}
 \label{thermal-expansion}
\end{figure}

In our calculations, we have used the lattice constant obtained at zero-temperature. For higher temperatures, the lattice constant increases due to thermal expansion. The latter is proportional to the Gruneisen parameter and elastic compliance of the material. Fig. \ref{Gvelocity} shows the mode Gruneisen parameters which are between 0 and 1.5, indicating that this  system is quite harmonic.
The calculated thermal expansion coefficients are shown in Fig.\ref{thermal-expansion}. The calculated room temperature thermal expansion coefficients of hexagonal BC$_{2}$N are $1.38\times10^{-6}$ and $1.60\times10^{-6}$, which are more than that of diamond,$1.06\times10^{-6}$\cite{mounet2005first,broido2013ab,stoupin2011ultraprecise}, and c-BN,$1.15\times10^{-6}$, \cite{slack1975thermal}, and closer to that of Si,$2.6\times10^{-6}$, \cite{okada1984precise,watanabe2004linear,broido2013ab}. The comparison with diamond and c-BN suggests that BC$_{2}$N may be a better candidate for thermal management in Si-based devices than diamond. 

\subsection{Thermal conductivity of hexagonal BC$_{2}$N}
\begin{figure}[pos=h]
  \includegraphics[height=6 cm]{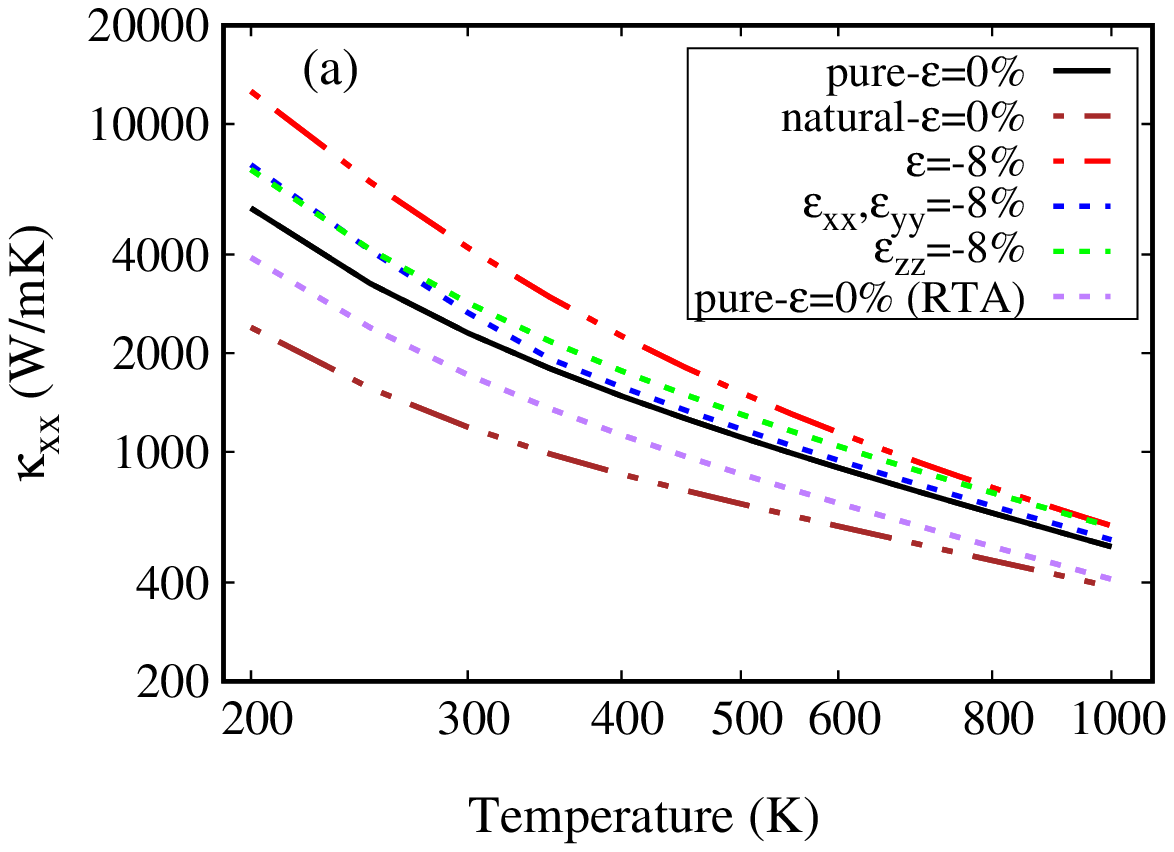}
   \includegraphics[height=6 cm]{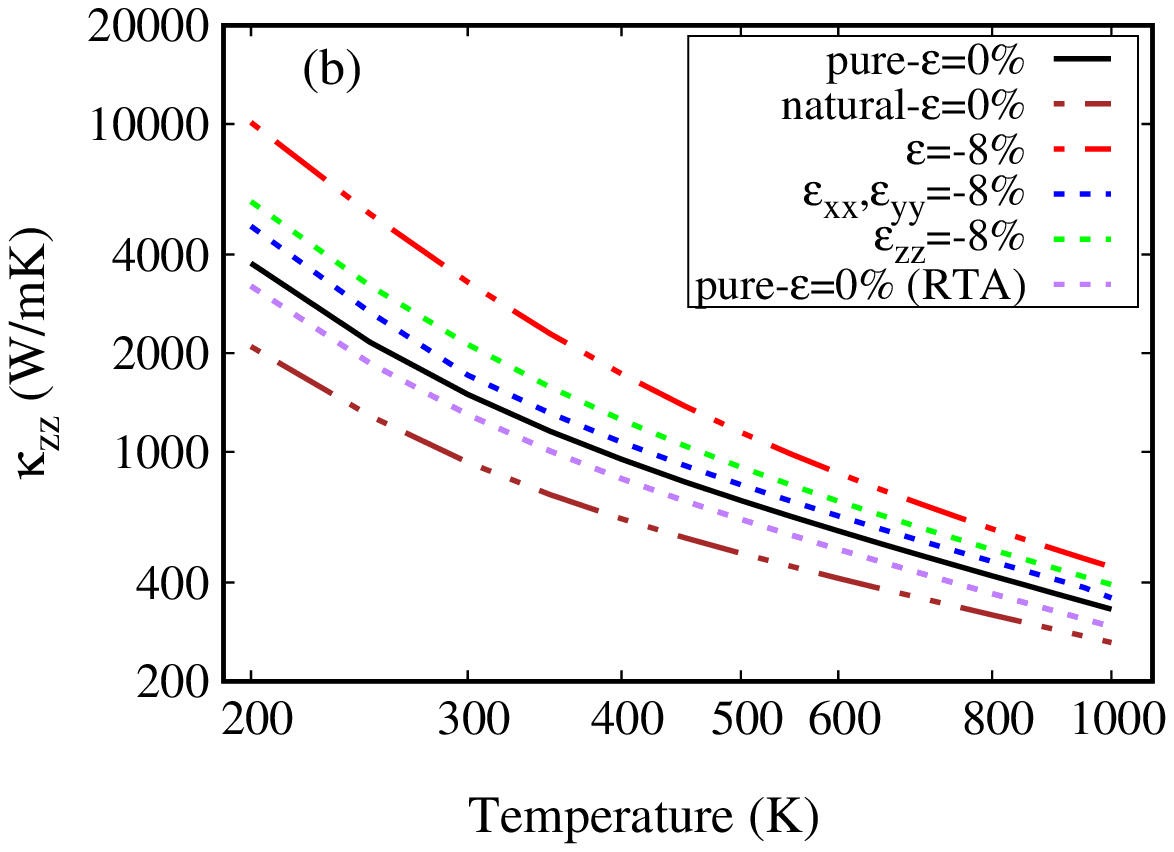}
    \caption{Thermal conductivity versus temperature under different compressive stresses: (a) in-plane direction $\kappa_{xx}$, and (b) cross plane direction $\kappa_{zz}$.
    }
  \label{thermal-conductivity}
\end{figure}

In the equilibrium geometry, we have investigated the thermal conductivity of both naturally occurring  and hypothetical isotopically pure hexagonal BC$_{2}$N. Fig.\ref {thermal-conductivity}(a,b) show the temperature dependence of calculated thermal conductivity along both directions from 200 to 1000 K. The results show that at room temperature, $\kappa_{nat}$= 1190 W/m-K, while $\kappa_{pure}$= 2305 W/m-K,   about 48\% larger than $\kappa_{nat}$ due to removal of the phonon isotope scattering, which is quite large at lower temperatures.  These results were obtained by considering the interaction cutoff for cubic IFCs calculations up to the seventh nearest neighbors. Table.\ref{thermal-conductivity-neighbors} shows these results in comparison with those resulting from including up to eleven nearest neighbors in the calculation of cubic force constants. Table \ref{thermal-conductivity-neighbors} shows that the converged results are around 10\% lower than thermal conductivity obtained from 7 nearest neighboring interaction, which is invisible in log scale of Fig.\ref{thermal-conductivity}.   

\begin{table}[pos=H]
\centering
\caption{Thermal conductivity at room temperature versus cumulative number of neighbors included in the calculation of cubic force constants}
\begin{tabular}{p{2.5cm} p{2.5cm} p{2.5cm}}

\hline\hline                      
 \# of Neighbors  & $7^{th}$ & $11^{th}$ \\[0.5ex]
\hline
$\kappa_{pure}$  & 2305  & 2073 (W/m-K) \\
$\kappa_{nat}$ & 1190 & 1083 (W/m-K) \\ 
\hline
\end{tabular}
\label{thermal-conductivity-neighbors}
\end{table}

The thermal conductivity of hexagonal BC$_{2}$N is evidently high due to the strong bond stiffness \cite{liu2018hexagonal} and light mass of atoms B-C-N, which we have shown to produce extremely high phonon frequencies as well as high acoustic velocities. 
There is an inverse relationship between the 
acoustic velocity scale and the phase space for phonon-phonon scattering\cite{ward2009ab}. 
The high phonon-frequency scale thus reduces the phase space for three-phonon scatterings, providing another justification for the high thermal conductivity of this material. 

In Fig.\ref{thermal-conductivity}, the calculated pure lattice thermal conductivity of hexagonal BC$_{2}$N within both relaxation time approximation (RTA) and the full iterated solution to the phonon Boltzmann equation are plotted as a function of temperature. It is evident from the figure that the RTA solution gives a poor approximation to the lattice thermal conductivity of hexagonal BC$_{2}$N. For example, at 300 K, the full iterated thermal conductivity is about $34\%$ higher than the result within RTA. This implies that the behavior of thermal conductivity is dominated by the normal scattering processes. This dominance manifests itself through the converged solution of the phonon BTE giving the thermal conductivity that lies higher than the RTA result. This can be compared to the case of diamond where the full thermal conductivity was 50\% larger than its RTA value.\cite{ward2009ab}. 

 \begin{figure}[pos=h]
 \centering
  \includegraphics[height=6 cm]{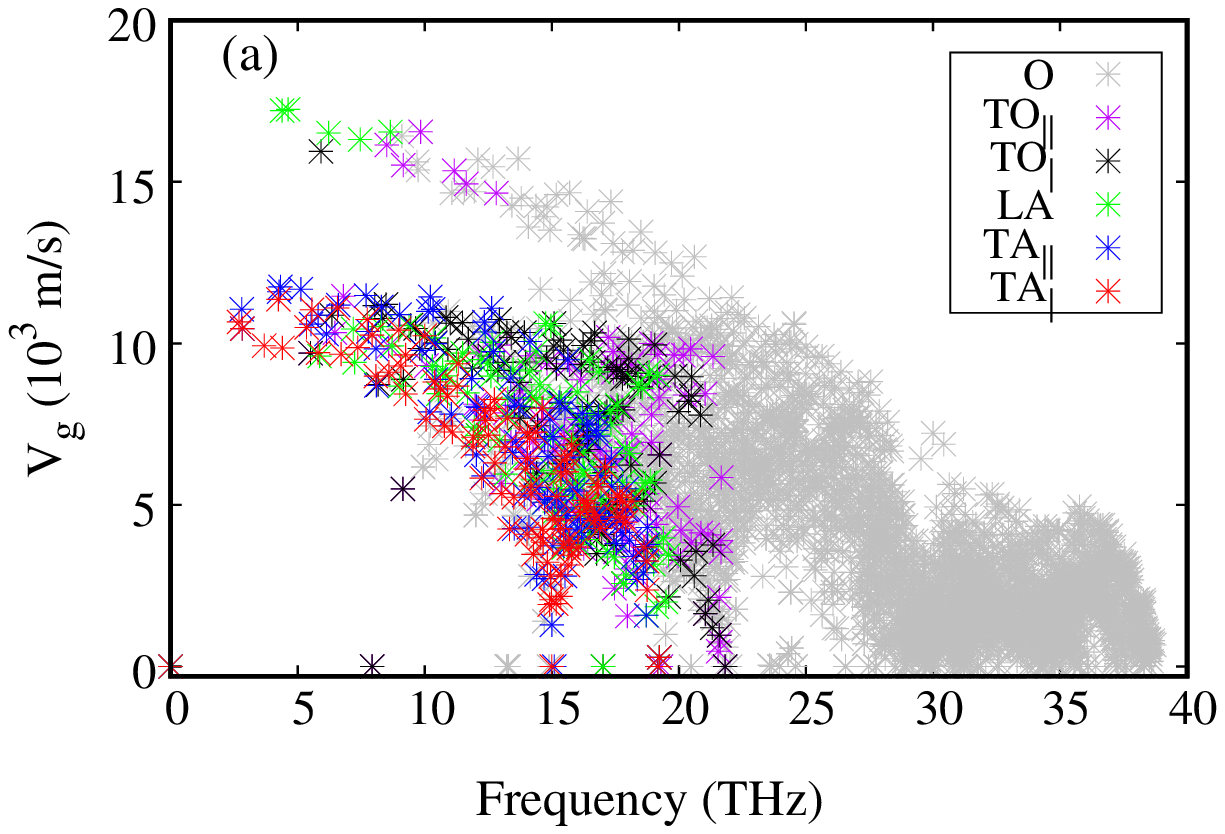}
  \includegraphics[height=6 cm]{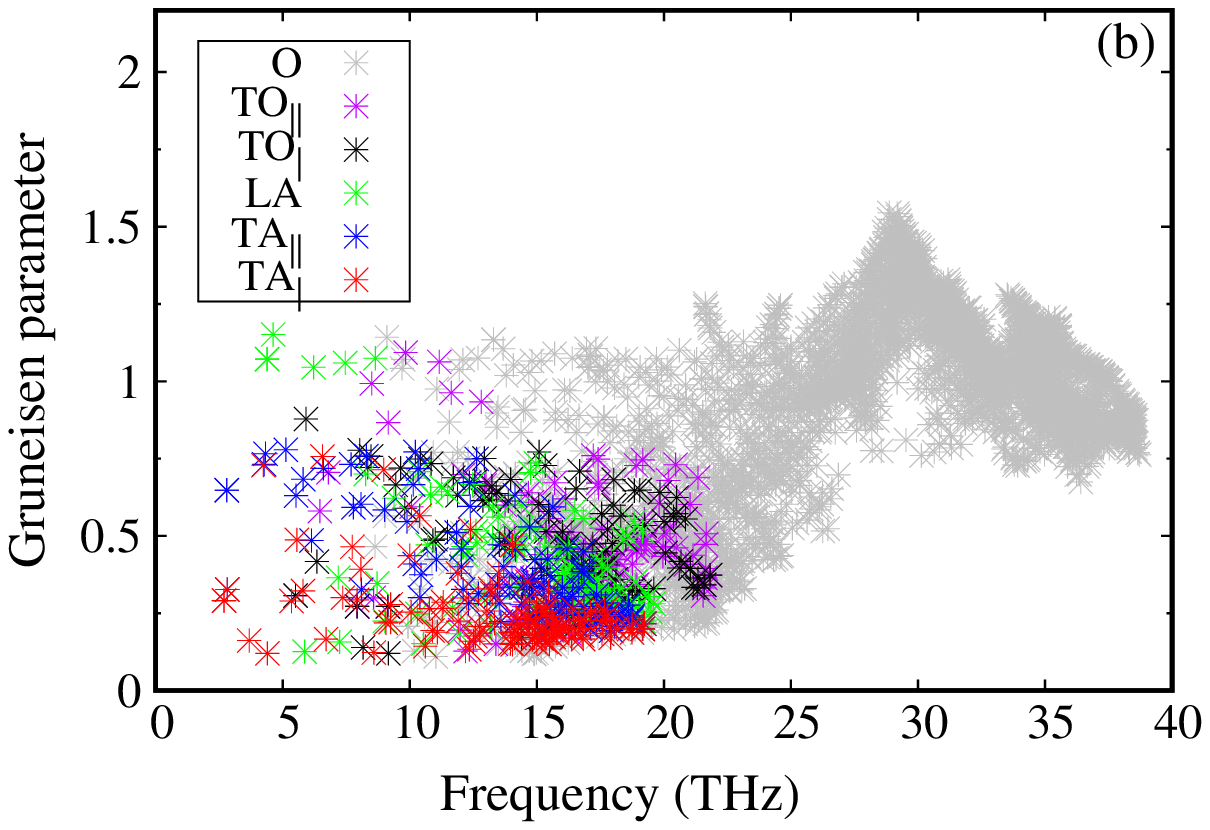}
   \caption{Phonon group velocities (a) and Gruneisen parameters (b) of each phonon mode as a function of phonon frequency for the equilibrium geometry of hexagonal BC$_{2}$N. Letter "O" represents the other optical modes. 
   }
  \label{Gvelocity}
\end{figure}

\begin{figure}[pos=h]
       \includegraphics[height=6 cm]{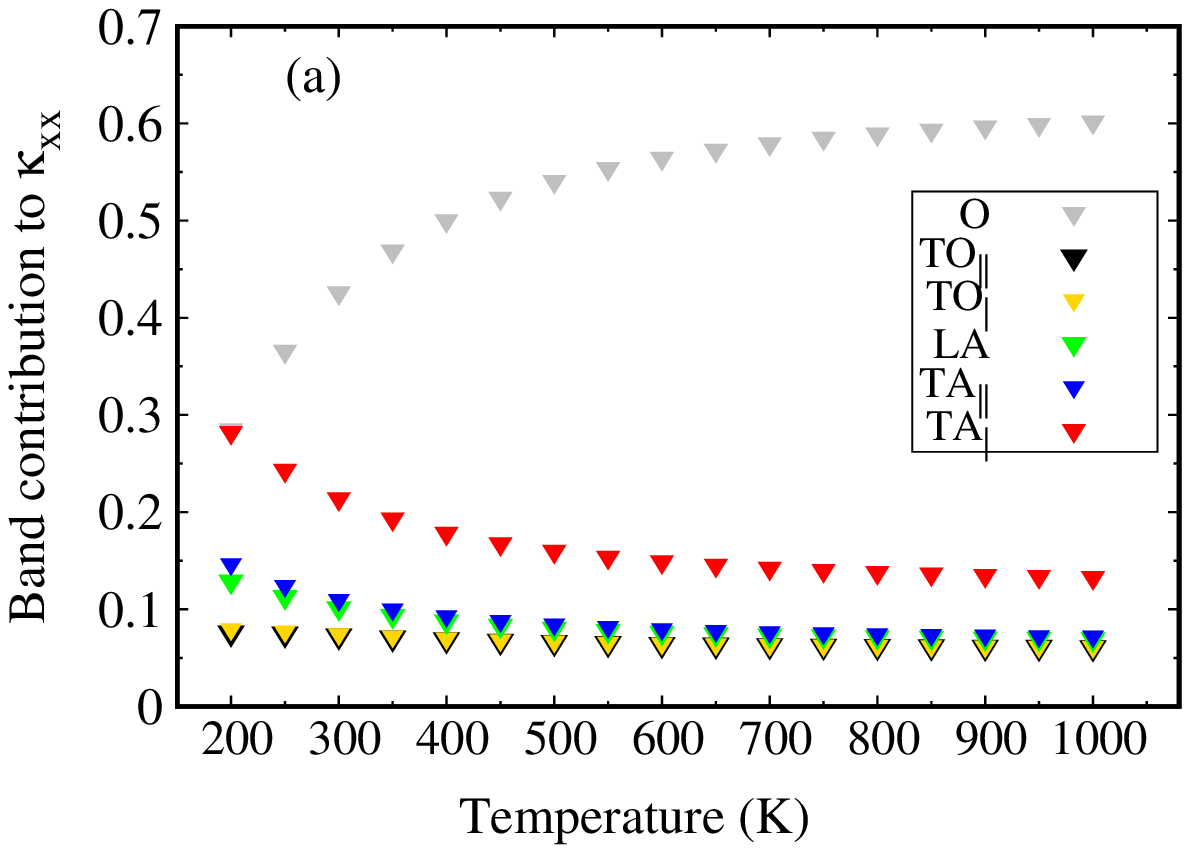}
   \includegraphics[height=6 cm]{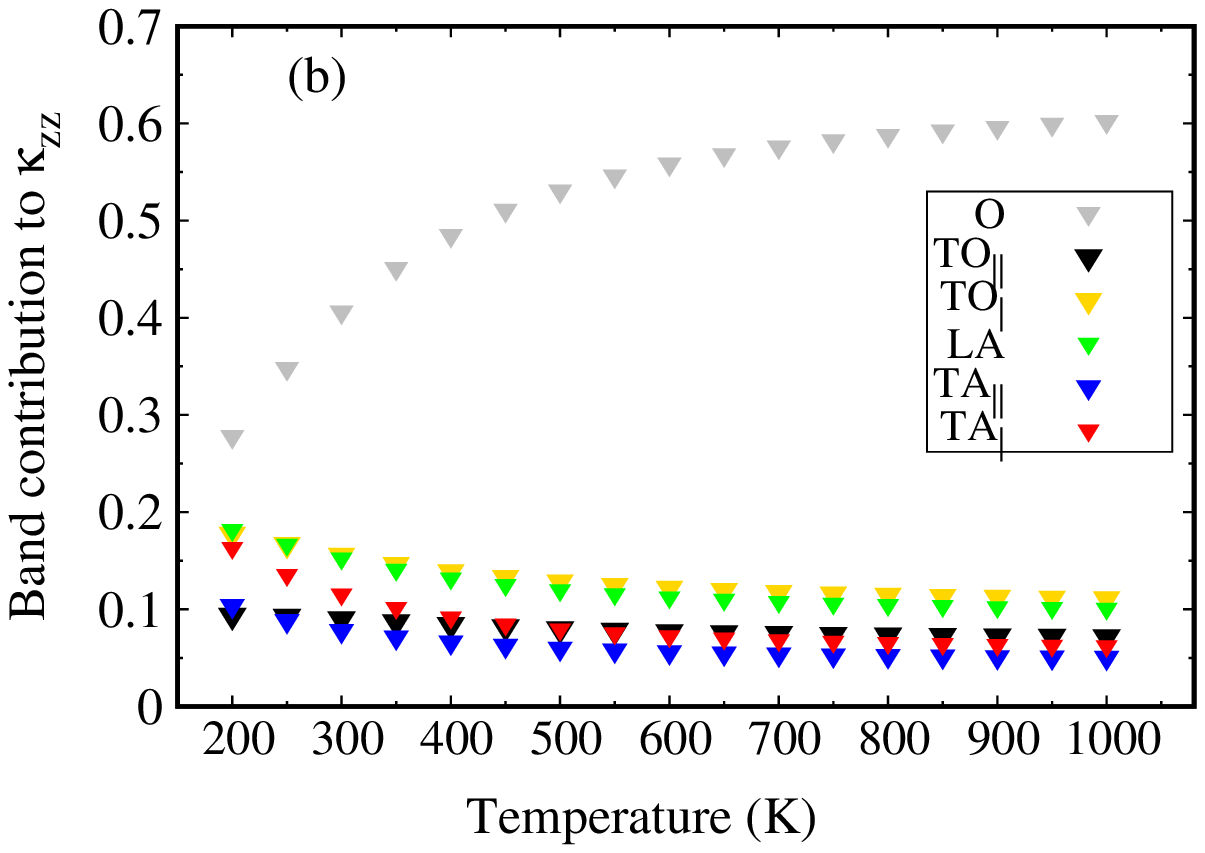}
    \caption{Normalized contribution of each phonon branch to the total thermal conductivity as a function of temperature for zero pressure in (a) in-plane $\kappa_{xx}$ and (b) cross-plane direction $\kappa_{zz}$. Letter "O" represents the other optical modes.}
  \label{mode-contributions}
\end{figure}

In most calculations of the lattice thermal conductivity, acoustic phonons are usually considered as the majority of heat-carrying phonons, and optical phonons provide essential scattering channels for the former\cite{ward2010intrinsic}. This is because optical phonons have small group velocities, so they have little direct contribution to the thermal conductivity, except in the nanoscale regime \cite{tian2011importance} where the long mean free paths of acoustic phonons become strongly reduced due to the small system size. In the equilibrium geometry of hexagonal BC$_{2}$N, as can be seen in  Fig.\ref{phonon-dispersion} and Fig.\ref{Gvelocity}, there are optical phonon modes with angular frequency around 20 THz with group velocities similar to that of the acoustic modes, which makes their contribution comparable to that of acoustic phonons. In order to illustrate this, we have calculated the contributions of different phonon branches to the lattice thermal conductivity of BC$_{2}$N, as shown in Fig.\ref{mode-contributions}(a,b). 

It can be seen that the contribution of TO1 and TO2 optic modes are roughly more than 8 percent in in-plane direction at room temperature, while the contribution of TA1 acoustic mode is 22 percent. Along the cross-plane direction, as shown in Fig.\ref{mode-contributions} (b), the contributions of TO1 and TO2 optic modes are more than that of LA and TA2 respectively.  It is also worth mentioning, as can be seen in table.\ref{compareing-gv-freq}, that the calculated maximum phonon frequencies of LO, transverse TA and longitudinal acoustic LA velocities of BC$_{2}$N are close to those of diamond \cite{ward2009ab} and c-BN\cite{mukhopadhyay2014polar}.

\subsection{The effect of pressure on thermal conductivity of hexagonal BC$_{2}$N}
Fig.\ref{thermal-conductivity}(a,b) shows the temperature dependence of the calculated thermal conductivity for hexagonal BC$_{2}$N from 200 to 1000 K under different compressive strains along the $x$ and $z$ directions, respectively. With increasing pressure, $\kappa_{pure}$ becomes larger than the corresponding P = 0 values throughout the range of temperatures. At compressive hydrostatic pressure of 8\%, P=150Gpa, $\kappa_{pure}^{xx}$ = 4196 W/m-K corresponding to nearly twofold increase compared to the P = 0 value. This ratio is comparable to that of diamond and c-BN, as can be seen in table\ref{thermal-conductivity-pressure}. These values are higher than those in any known material at ambient pressure and temperature. It can also be seen from blue and green lines in Fig.\ref{thermal-conductivity}(a,b) that thermal conductivity at room temperature can be increased to 2659 W/mK,1715W/m-K and 2848 W/mk, 2127W/m-K in both in-plane and cross-plane directions, by applying compressive pressure only laterally or vertically, respectively. 

\begin{table}[pos=h]
\centering
\caption{Zone center frequencies, transverse and longitudinal acoustic velocities, and thermal conductivity of hexagonal BC$_{2}$N, c-BN, and diamond (experimental values appear in parentheses). }

\begin{tabular}{p{1.8cm} p{1.2cm} p{1.5cm} p{1.8cm} p{1.1cm}}

\hline\hline                      
   & \tblmultiline{$\omega_{LO}$ \\ (THz)} & \tblmultiline{$v_{TA}$ \\ (m/s)} & \tblmultiline{$v_{LA}$ \\ (m/s)} & \tblmultiline{$\kappa$ \\ (W/m-K)} \\[0.5ex]
\hline

BC$_{2}$N & 38.33 &\tblmultiline{11362 \\ 11748} &17237 & 2305  \\
\vspace{0.1cm}
c-BN\cite{mukhopadhyay2014polar} &  \vspace{0.1cm}38.49 & \vspace{0.1cm} 11220 &  \vspace{0.1cm}15110 & \vspace{0.1cm} 2145\cite{mukhopadhyay2014polar}\\ 
 &   & (11800\cite{wang2004determination}) &  (15410\cite{wang2004determination}) & \\ 
\vspace{0.1cm}
diamond \cite{ward2009ab} & \vspace{0.1cm}39.17 & \vspace{0.1cm}12567[100]&\vspace{0.1cm} 17326[100] & \vspace{0.1cm} 3450 \cite{broido2012thermal} \\
 & (40.11\cite{warren1967lattice}) & (12830\cite{mcskimin1972elastic})&\ (17520 \cite{ward2009ab})  & \\

\hline
\end{tabular}
\label{compareing-gv-freq}
\end{table}

\begin{figure*}
\begin{multicols}{2}
    \includegraphics[height=6 cm,width=1.1\linewidth]{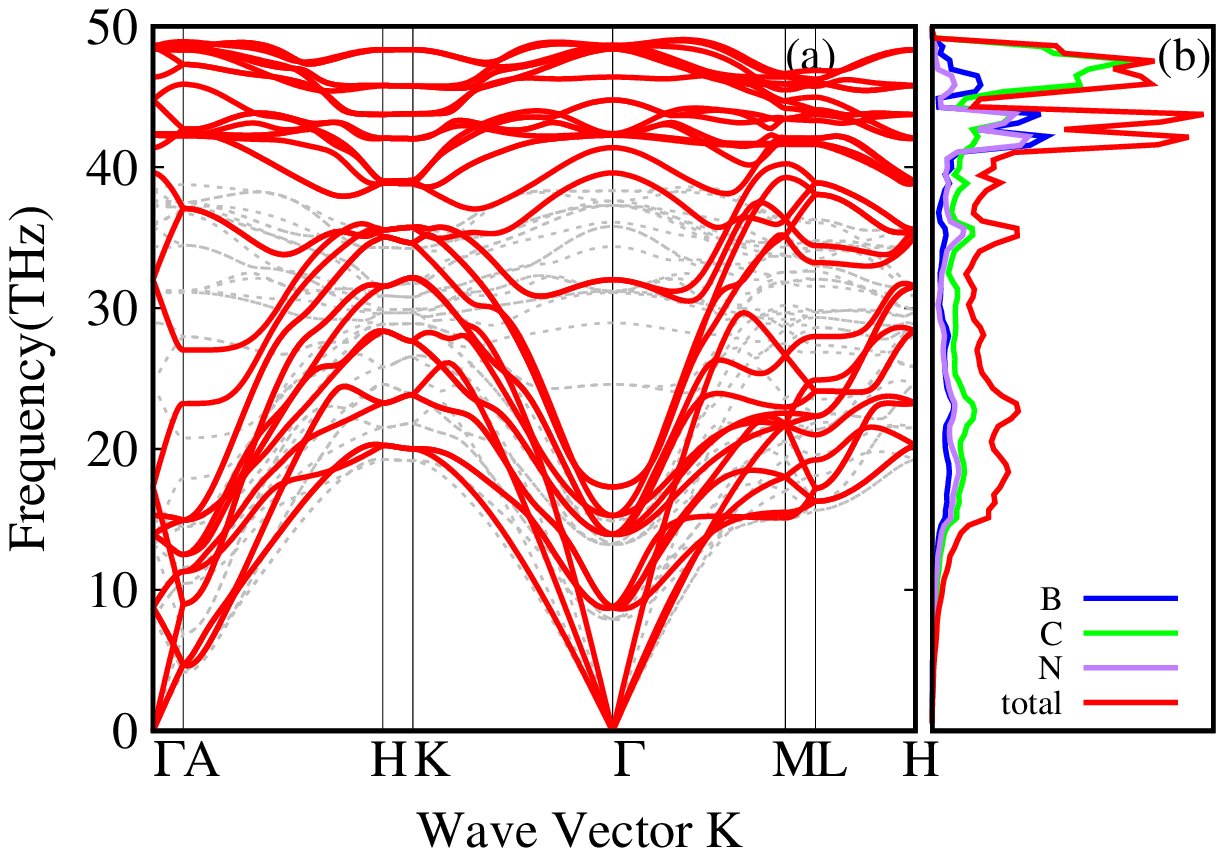}\par 
    \includegraphics[height=6 cm,width=1.1\linewidth]{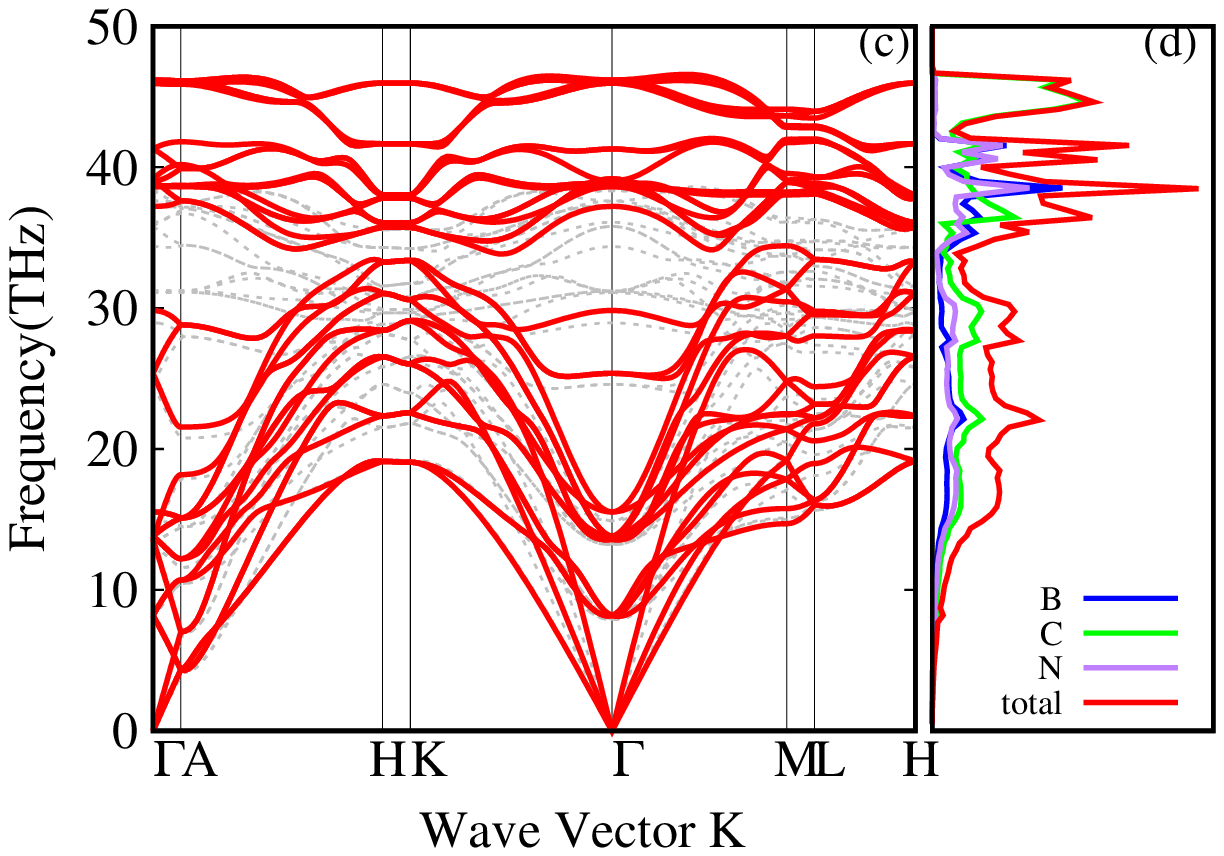}\par
    \end{multicols}
\begin{multicols}{2}
    \includegraphics[height=6 cm,width=1.1\linewidth]{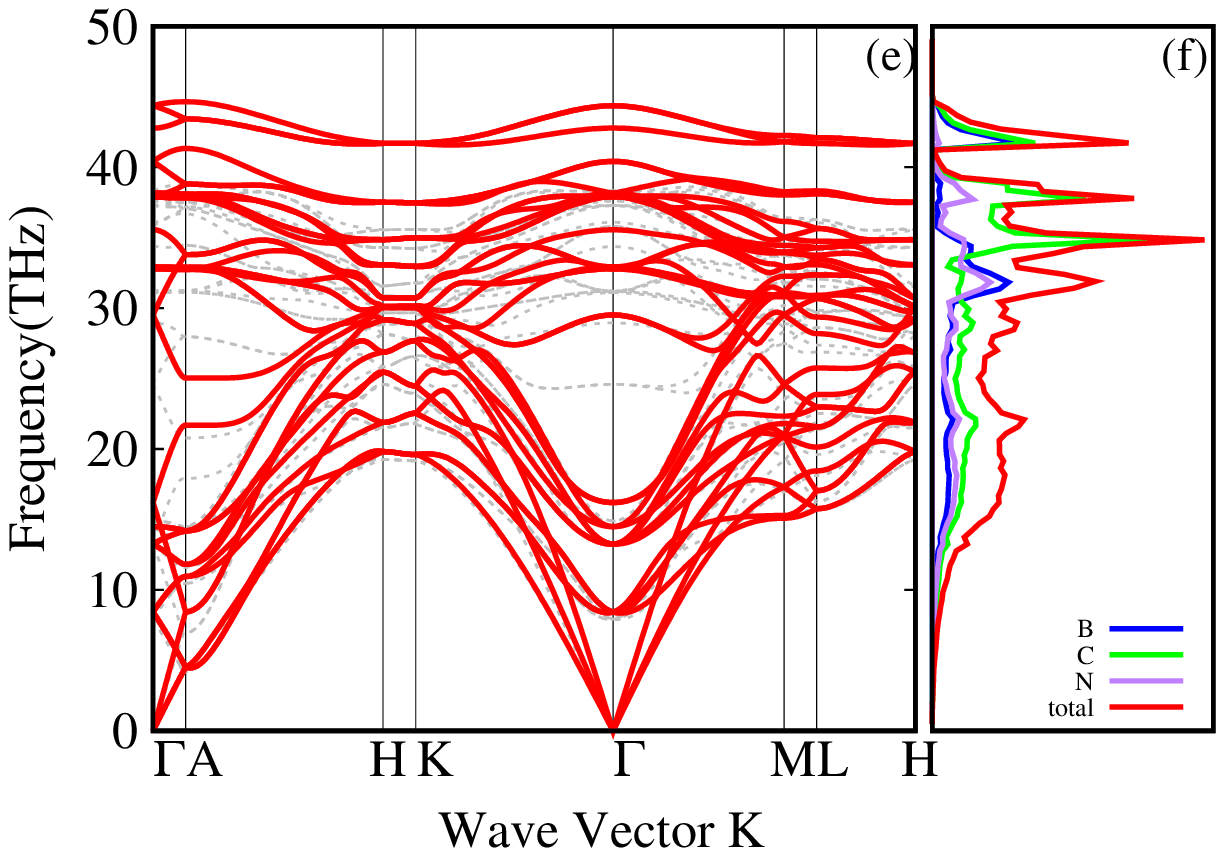}\par
    \includegraphics[height=5.5 cm,width=1.1\linewidth]{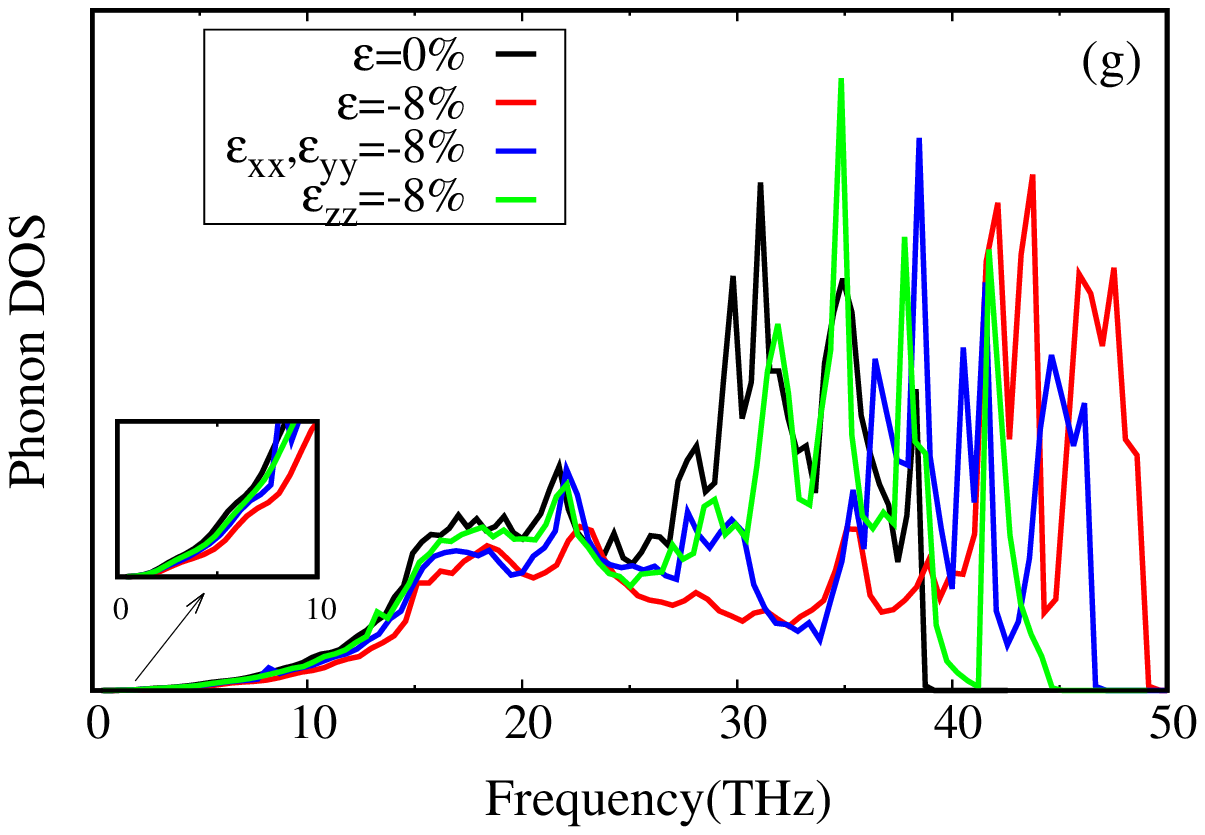}\par
\end{multicols}
\caption{Phonon dispersion curves and density of  states of hexagonal BC$_{2}$N under compressive hydrostatic pressure $\varepsilon=8\%$ (a,b), in-plane compressive pressure $\varepsilon_{xx}=\varepsilon_{yy}=8\%$  (c,d), and vertical compressive pressure $\varepsilon_{zz}=8\%$  (e,f). (For comparison, the dispersion at equilibrium geometry is  plotted with gray curves), (g) The comparison of phonon density of states of hexagonal BC$_{2}$N under different compressive pressures.}
  \label{phonons-DOS}
\end{figure*}

To understand how pressure affects the thermal conductivity of hexagonal BC$_{2}$N, we examined phonon dispersions and the corresponding phonon density of states at various pressures. These are shown in Fig.\ref{phonons-DOS}. As can be seen in these figures, the frequencies of all phonon modes are positive, attesting to the dynamical stability of all compressed hexagonal BC$_{2}$N. According to simulation results, all elastic constants of compressed hexagonal BC$_{2}$N satisfy the stability criteria. For clarity and ease of comparison, all of these phonon dispersions are superimposed on the equilibrium geometry's dispersion plotted with dashed gray lines. It can be seen that under pressure the bandwidth is increased. This has two consequences: group velocity of acoustic and optical modes have to increase, leading to a higher thermal conductivity quadratic in that increase. The second effect which is less transparent, is the reduction in the scattering phase space. As can be seen in Eq.\ref{em-abs}, a rescaling of the frequencies by a factor of $s$ leads to a rescaling of the phase space integral by a factor of $s^{-5}$ if we take the classical limit of the distribution function $f_{\lambda} \simeq kT/\hbar \omega_{\lambda}$ or $s^{-4}e^{-\alpha s}$ if one is at low temperature. So, assuming a simple rescaling of the frequencies under applied pressure, according to Eq.\ref{Klattice} there is a gain of $s^2$ from the group velocities and a gain of $s^4$ or $s^5$ from the scattering time. It is however not very clear how anharmonicity, or cubic force constants, scale with applied pressure. The Gruneisen parameter is a good indication of anharmonicity strength and will be discussed later. In reality, the scaling is not the same for all modes, and this argument should just be taken as a support for explaining the increase in $\kappa$ with pressure.


\begin{table}[pos=H]
\centering
\caption{Thermal conductivity under compressive hydrostatic pressure at room temperature in comparison with its value at P=0Gpa}

\begin{tabular}{ p {0.5cm} p{1.7cm}| p{2cm} | p{3cm} }

\hline\hline 
  & & $\kappa_{pure}$ (W/m-K) at P=0 Gpa & $\kappa_{pure}$ (W/m-K)  \\[0.5ex]
\hline
 & BC$_{2}N$ &  2305 & 4196   P=150 Gpa\\[0.5ex]
\hline
 & diamond\cite{broido2012thermal} &   3450  &  6880   P=125 Gpa\\
\hline
 & c-BN\cite{mukhopadhyay2014polar} &  2113    (2145\cite{lindsay2013first})  &   3400  P=150 Gpa \\ 

\hline
\end{tabular}
\label{thermal-conductivity-pressure}
\end{table}

As we can see in Fig.\ref{phonons-DOS}, the scaling under hydrostatic pressure is nearly 25\% while that along the $z$ direction is only about 10\%. We can therefore expect the largest increase in the case of hydrostatic compression followed by in-plane and cross-plane compressions.

\begin{figure*}
\begin{multicols}{3}
    \includegraphics[height=6 cm,width=1.1\linewidth]{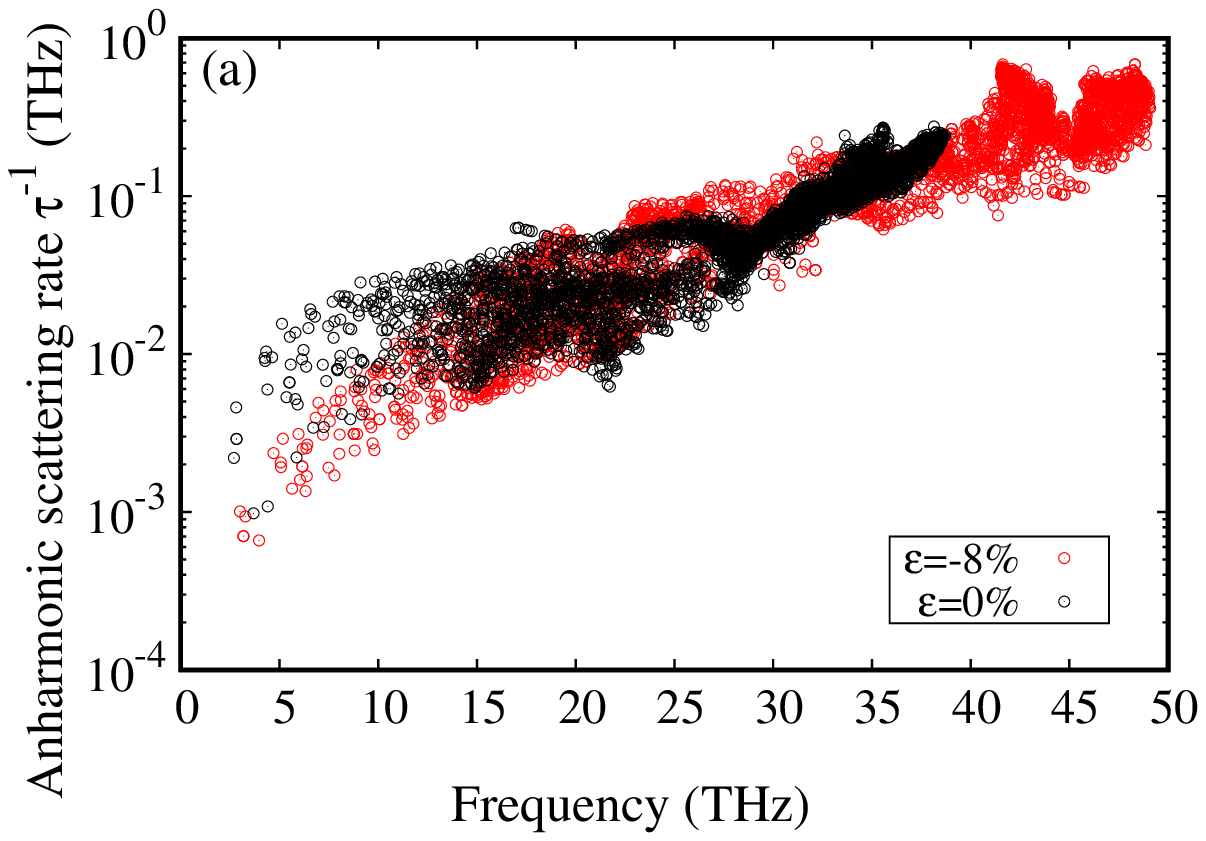}\par 
    \includegraphics[height=6 cm,width=1.1\linewidth]{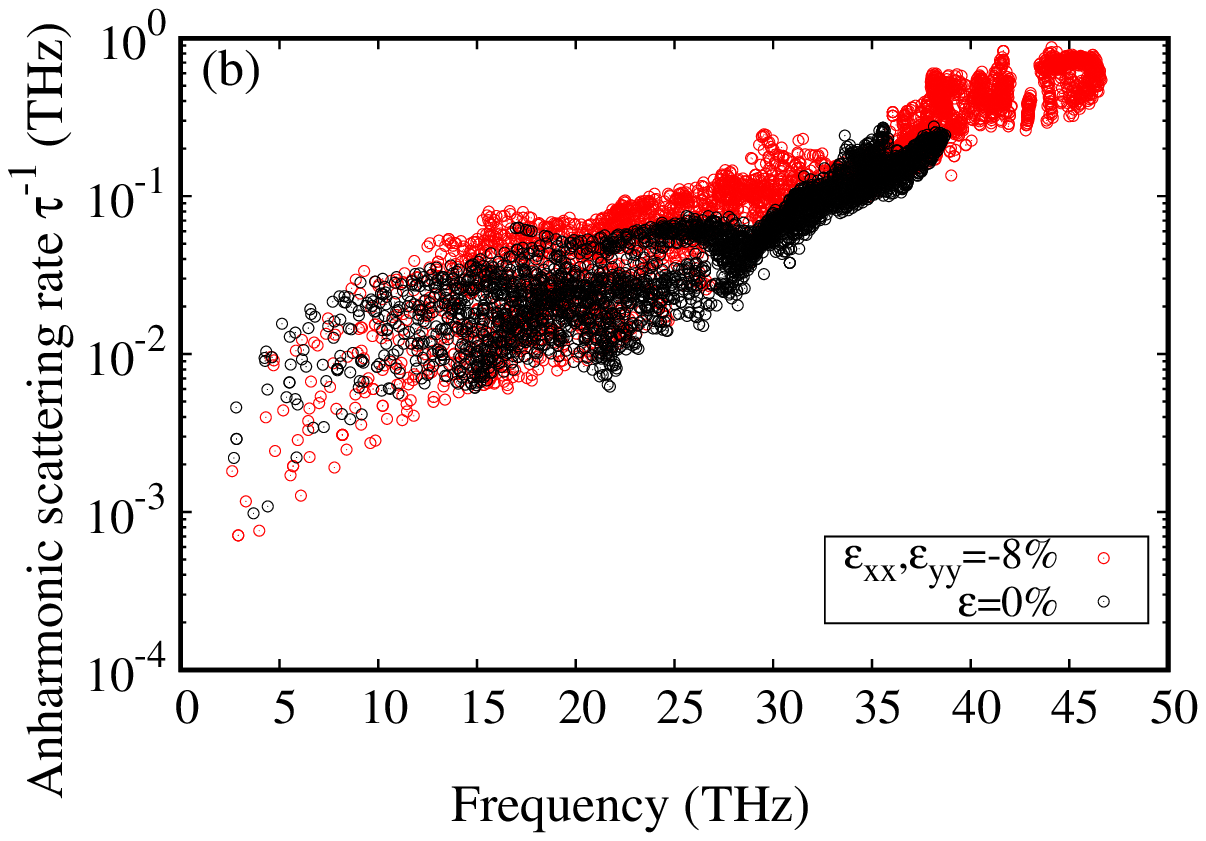}\par 
    \includegraphics[height=6 cm,width=1.1\linewidth]{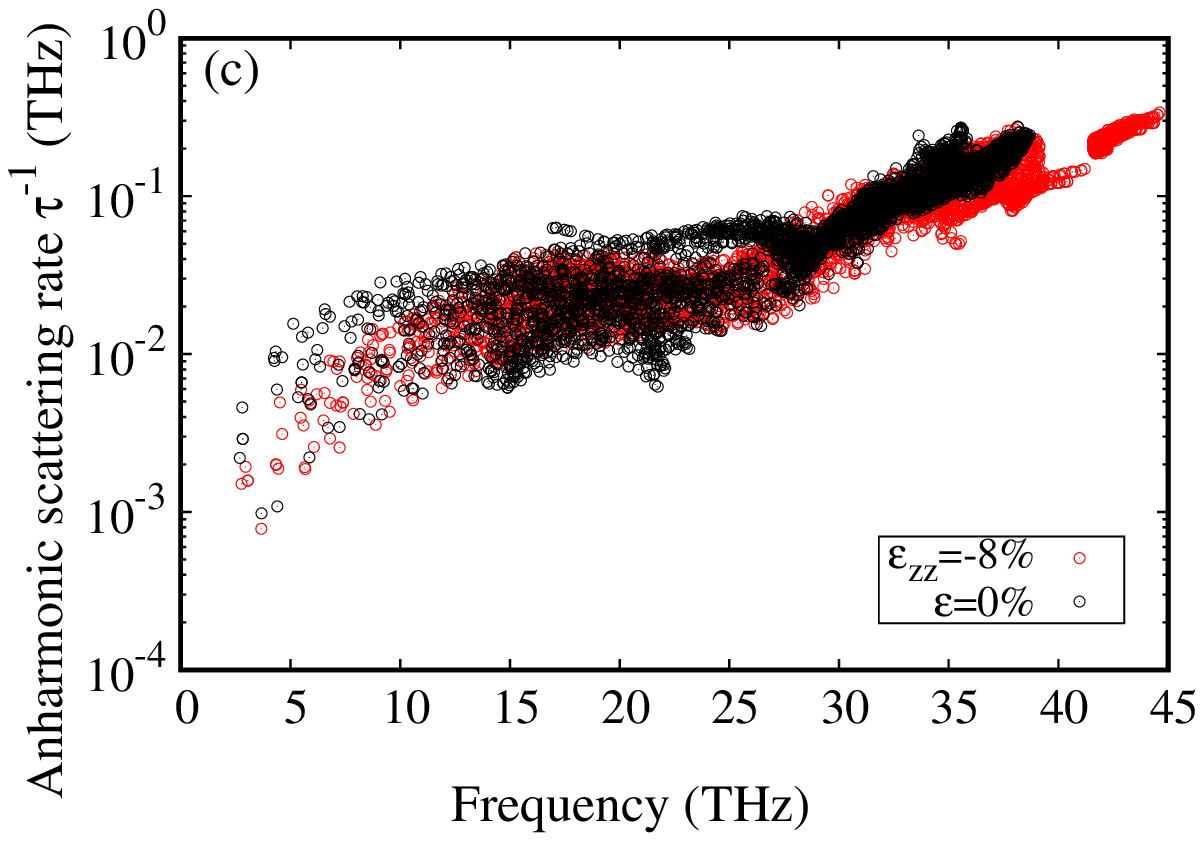}\par 
    \end{multicols}
\caption{ Anharmonic phonon scattering rates at 300 K compared to uncompressed, for hydrostatic (a), in-plane (b), and cross-plane (c) applied pressures. 
}
  \label{anharmonic-SRs}
\end{figure*}

\begin{figure}[pos=H]
\centering
  \includegraphics[height=5 cm]{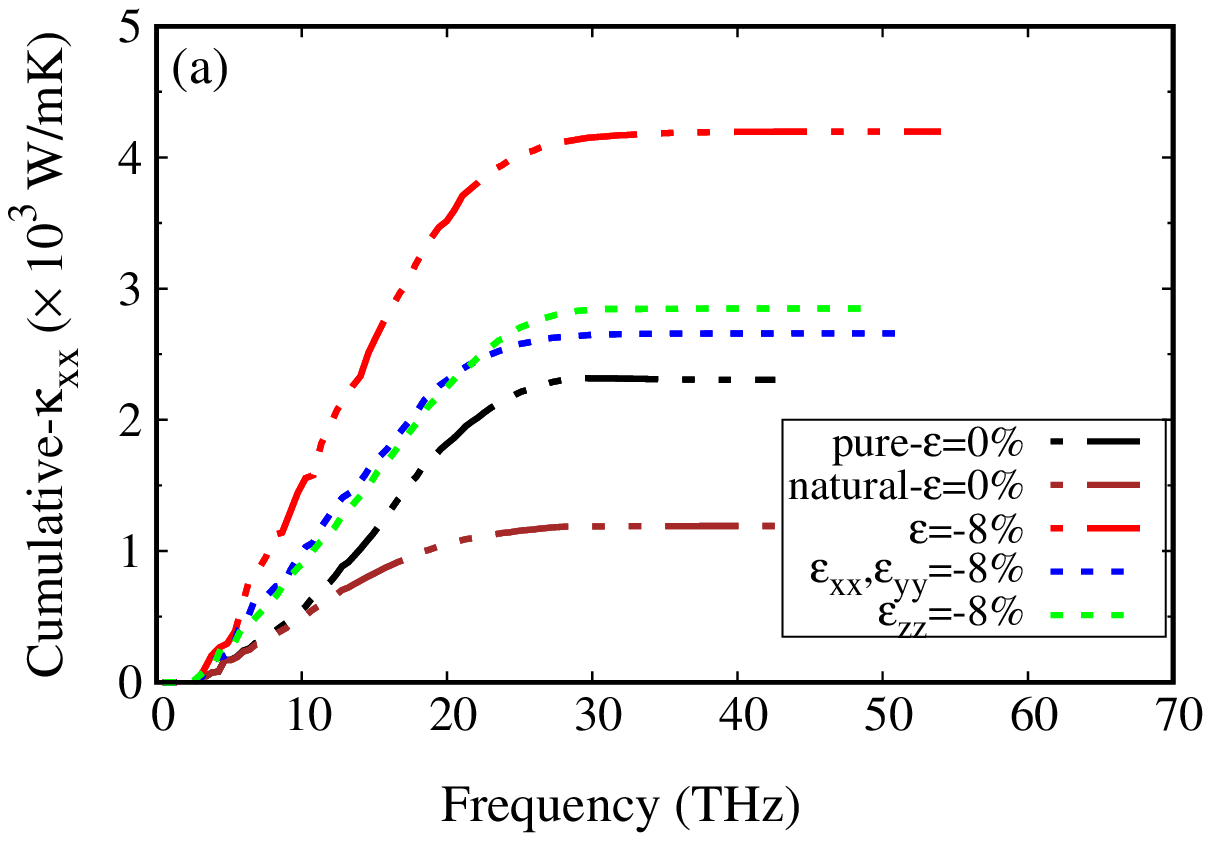}
  \includegraphics[height=5 cm]{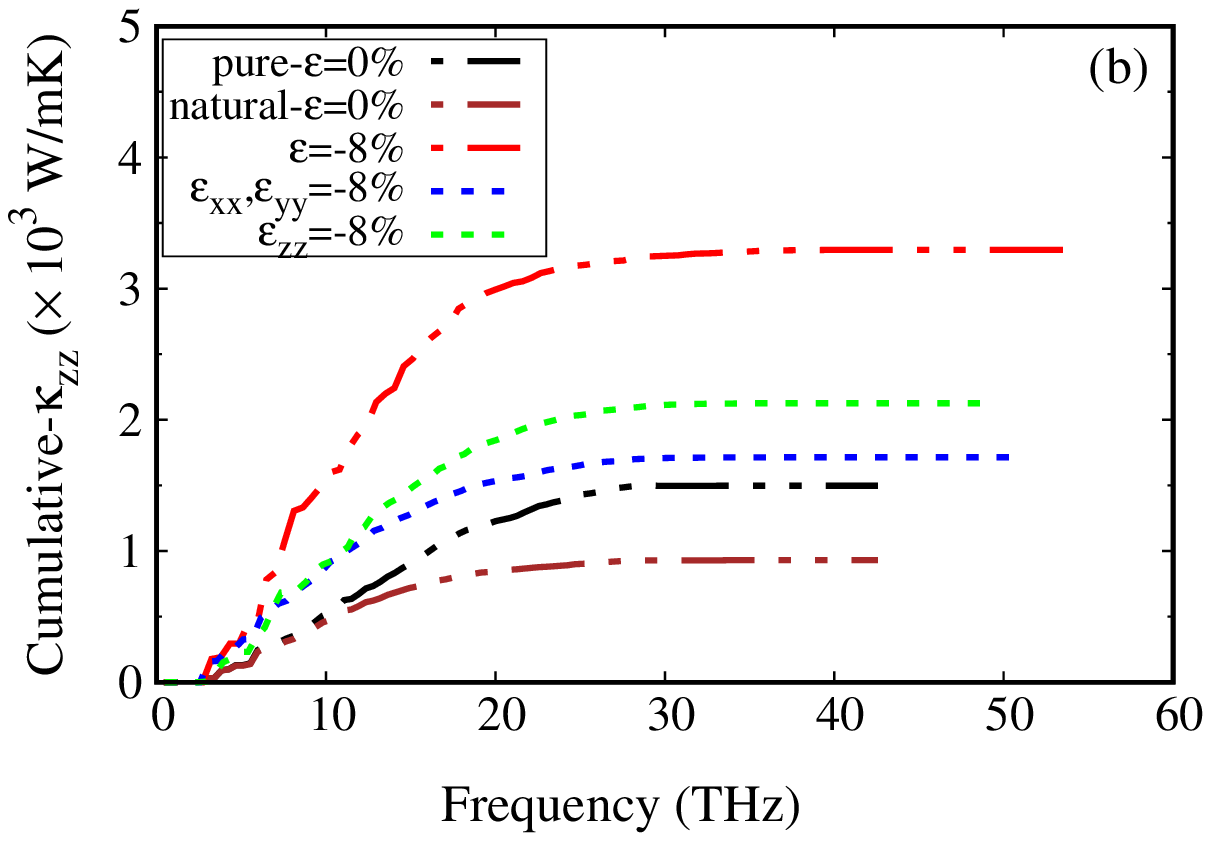}
    \caption{Cumulative thermal conductivity versus phonon frequency under the different pressures at T=300 K in (a) in-plane $\kappa_{xx}$, (b) cross plane directions$\kappa_{zz}$. }
  \label{cumulative}
\end{figure}

To highlight the connection between the optical phonon frequencies and thermal conductivity, the anharmonic scattering rates for the different pressures were calculated within the relaxation time approximation. The room temperature anharmonic scattering rates of hexagonal BC$_{2}$N under the different pressures are compared to the anharmonic scattering rates of the equilibrium geometry in Fig.\ref{anharmonic-SRs}. The cumulative plots of frequency dependent contributions to both components of  $\kappa_{RT}$, in Fig.\ref{cumulative}(a,b), show that thermal conductivity $\kappa_{RT}$ under each pressure is almost entirely due to phonons with angular frequency below 20 THz, corresponding to mean free paths of approximately 10 $\mu$m.  At compressive hydrostatic pressure of 8\%, phonons with $\omega$ up to 20 THz contribute more than 80\% to the thermal conductivity at room temperature. For this range of frequency, the scattering rates are seen to strongly decrease with pressure except for the in-plane case, as can be seen in Fig.\ref{anharmonic-SRs}(a-c).  

At compressive hydrostatic pressure 8\%, the scattering rates are smaller than half of that of the equilibrium geometry. 
It can be also seen that, at compressive hydrostatic pressure 8\%, the scattering rates has the lowest values, and the scattering rates from lateral compressive pressure are more than that of due to vertical compressive pressure.  

 \begin{figure}[pos=H]
\centering
  \includegraphics[height=5.5 cm]{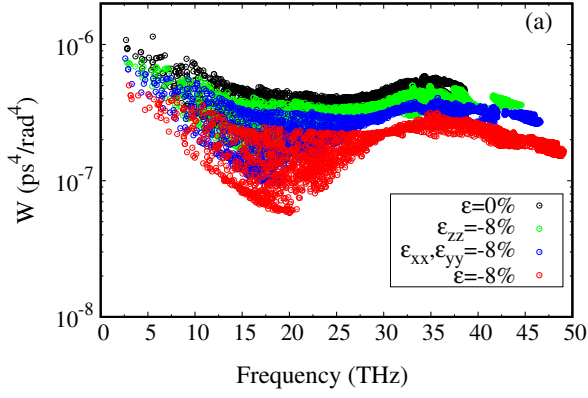}

    \caption{  Weighted phase space at room temperature vs frequency for different pressures.
    }
  \label{WP3}
\end{figure}
\begin{figure}[pos=H]
\centering
  \includegraphics[height=5 cm]{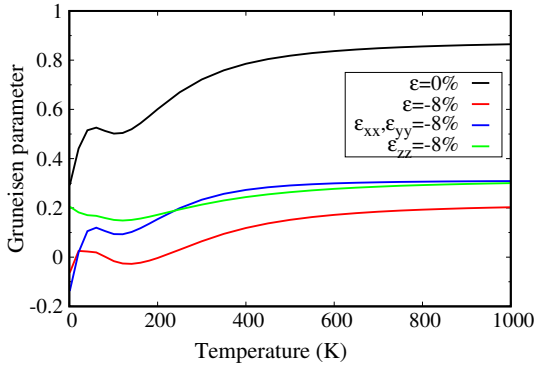}
    \caption{Gruneisen parameters as a function of temperature for hexagonal BC$_{2}$N under different compressive pressures. }
  \label{G_T}
\end{figure}

In Fig. \ref{WP3} and \ref{G_T}, we separate the role of anharmonicity strength and phase space in determining the change in the scattering rates. As can be seen in Fig.\ref{WP3},
and as discussed in the scaling argument, if the phonon bandwidth is increased, the scattering phase space will strongly decrease. This can be seen even in the logarithmic scale of Fig.\ref{WP3}. From Fig.\ref{G_T}, we can see that the Grunesien parameter, which is a measure of anharmonicity strength, has also strongly decreased by nearly four-fold with compressive pressure; the strongest suppression being for the hydrostatic pressure.

\begin{figure}[pos=H]
\centering
  \includegraphics[height=5 cm]{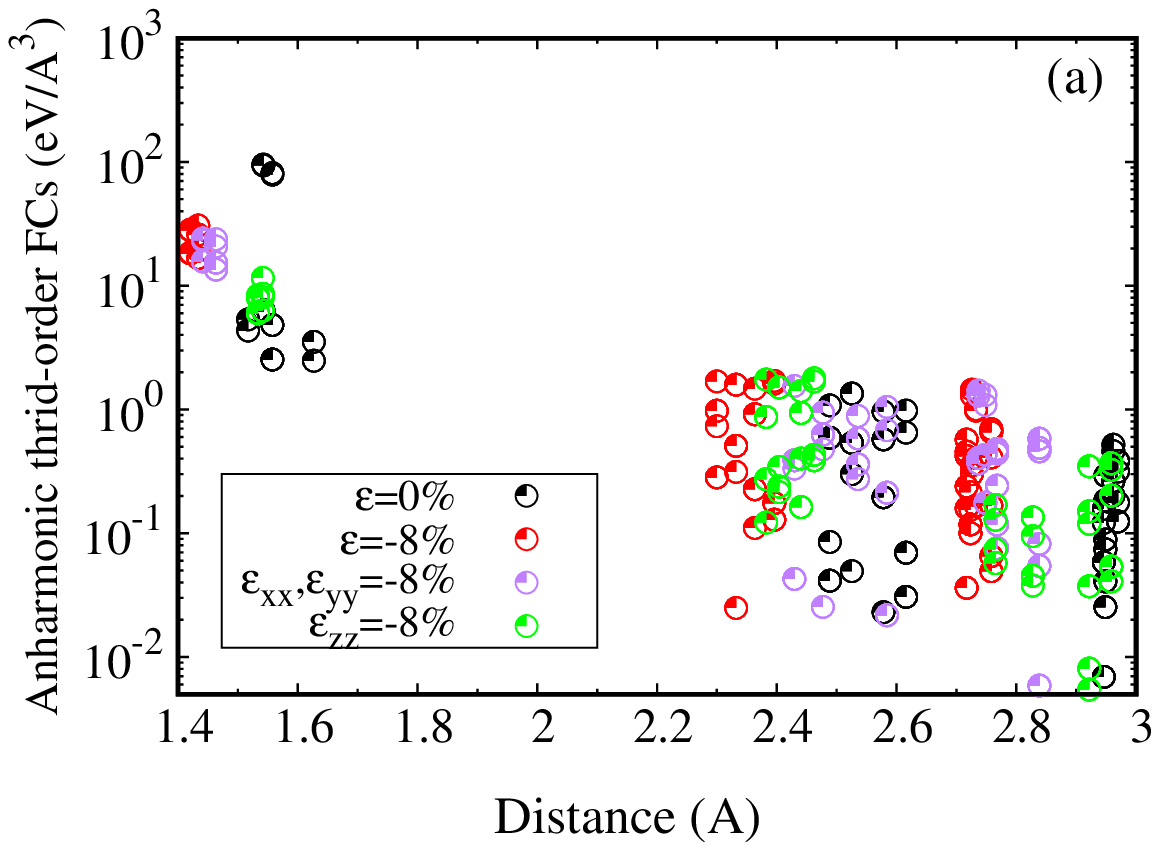}
  \includegraphics[height=5 cm]{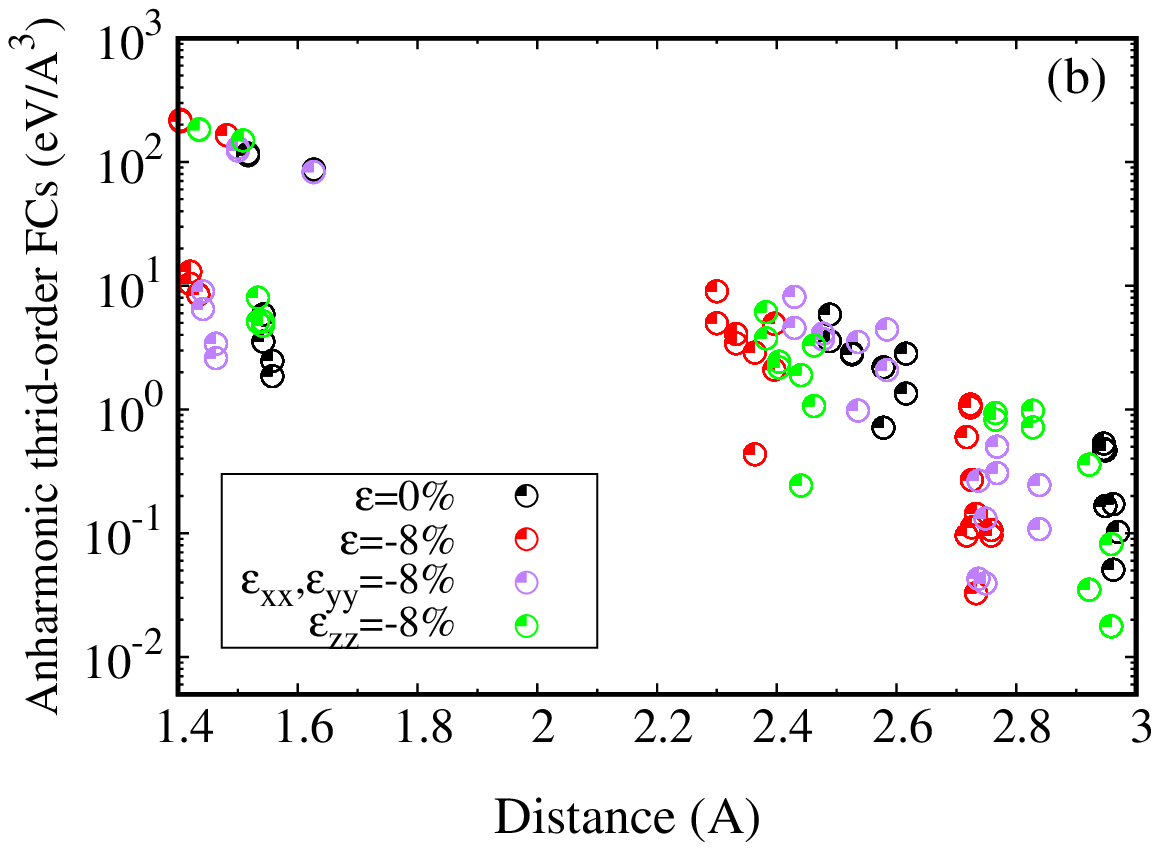}
    \caption{Largest anharmonic force constants along (a) yyy and (b) zzz versus distance}
   \label{FC3}
\end{figure}

To better understand this effect, we plot in Fig.\ref{FC3} the magnitude of cubic force constants as a function of bond length at various pressures. Of the 27 possible tensor elements, we have chosen the largest ones which happen to be the $yyy$ and $zzz$ components. It can be seen that IFCs along $yyy$ shown in Fig.\ref{FC3}(a) are generally smaller than the $zzz$ components shown in Fig.\ref{FC3}(b). The largest $yyy$ components decrease under pressure. Furthermore, the decrease in the Gruneisen parameter with pressure can be explained by noting that it is inversely proportional to the square of phonon frequencies which increase with pressure.



\begin{figure*}
\begin{multicols}{3}
    \includegraphics[height=6 cm,width=1.1\linewidth]{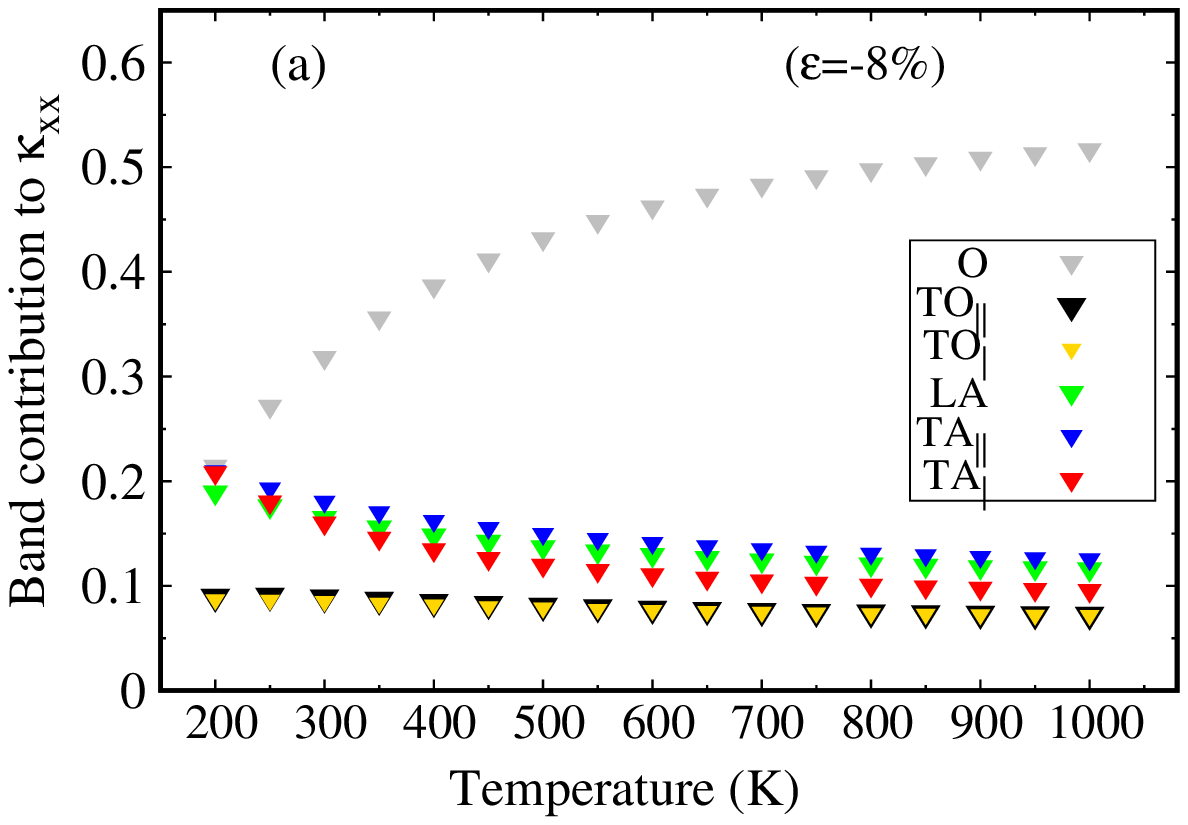}\par 
    \includegraphics[height=6 cm,width=1.1\linewidth]{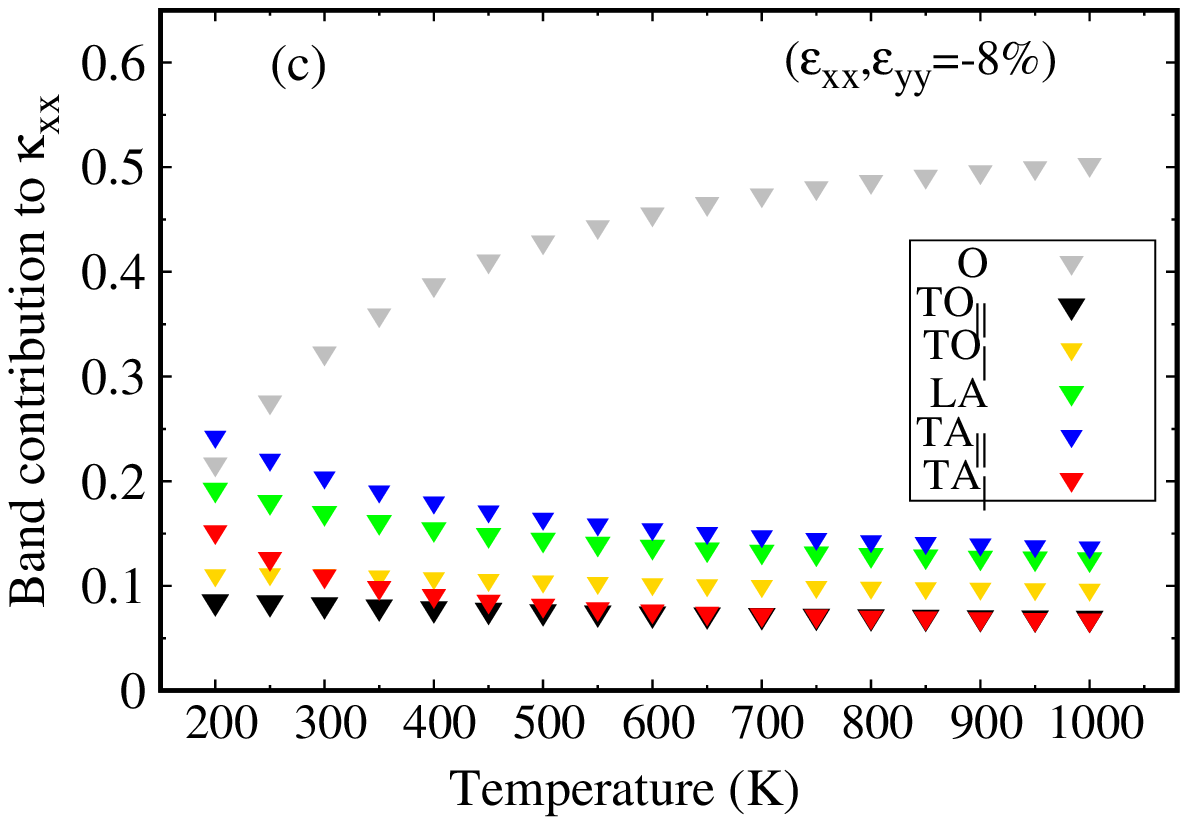}\par 
    \includegraphics[height=6 cm,width=1.1\linewidth]{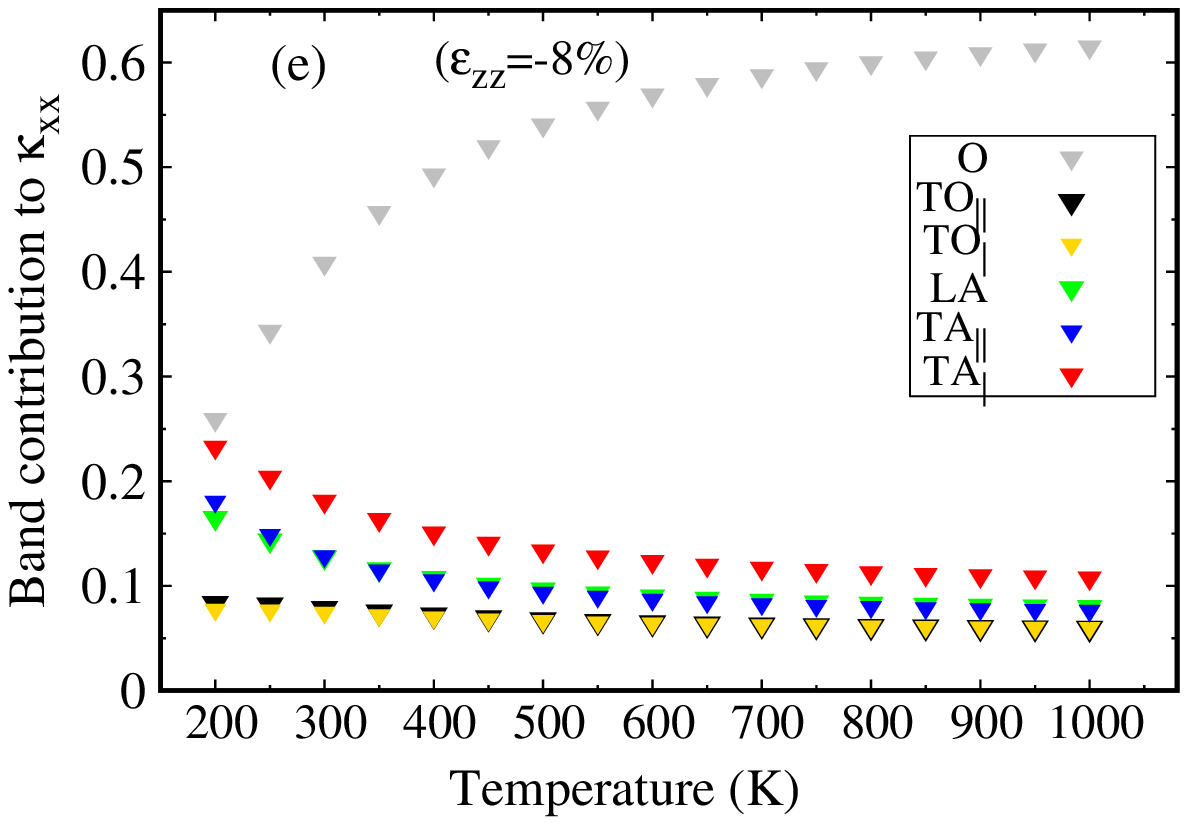}\par 
    \end{multicols}
\begin{multicols}{3}
    \includegraphics[height=6 cm,width=1.1\linewidth]{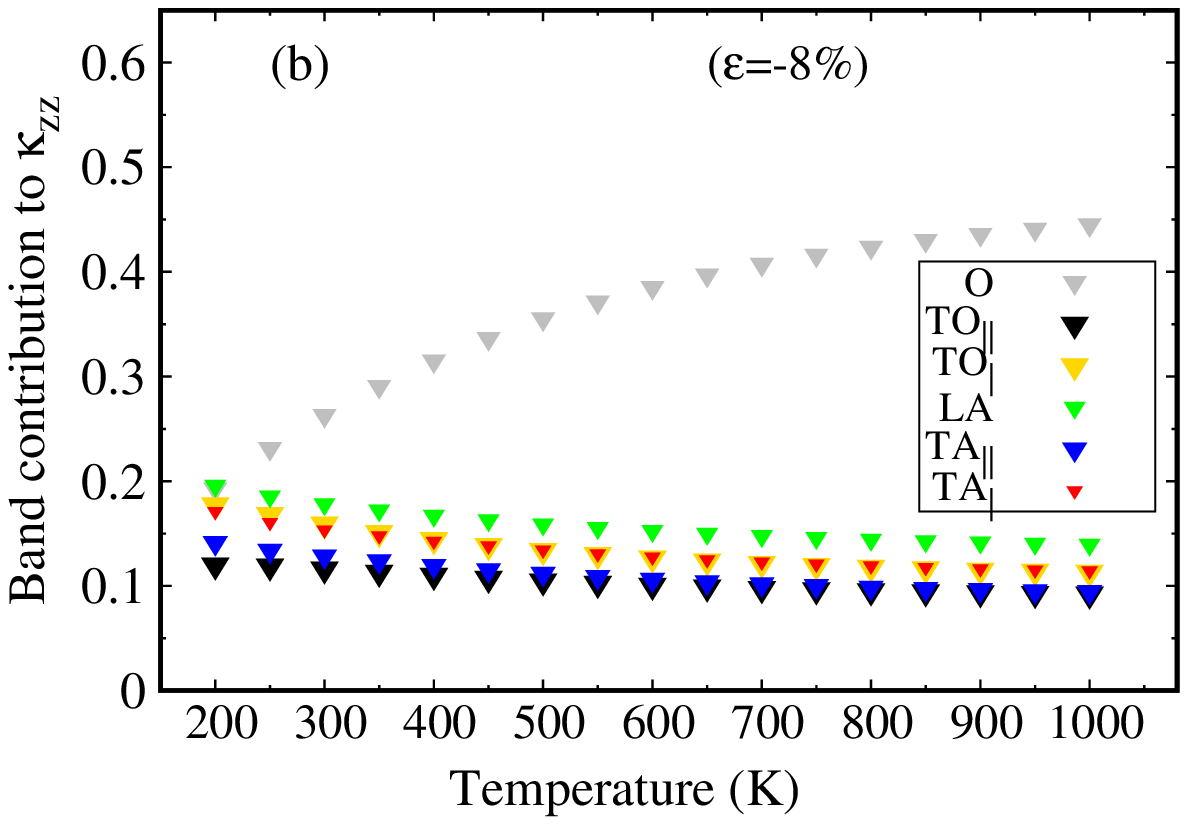}\par
    \includegraphics[height=6 cm,width=1.1\linewidth]{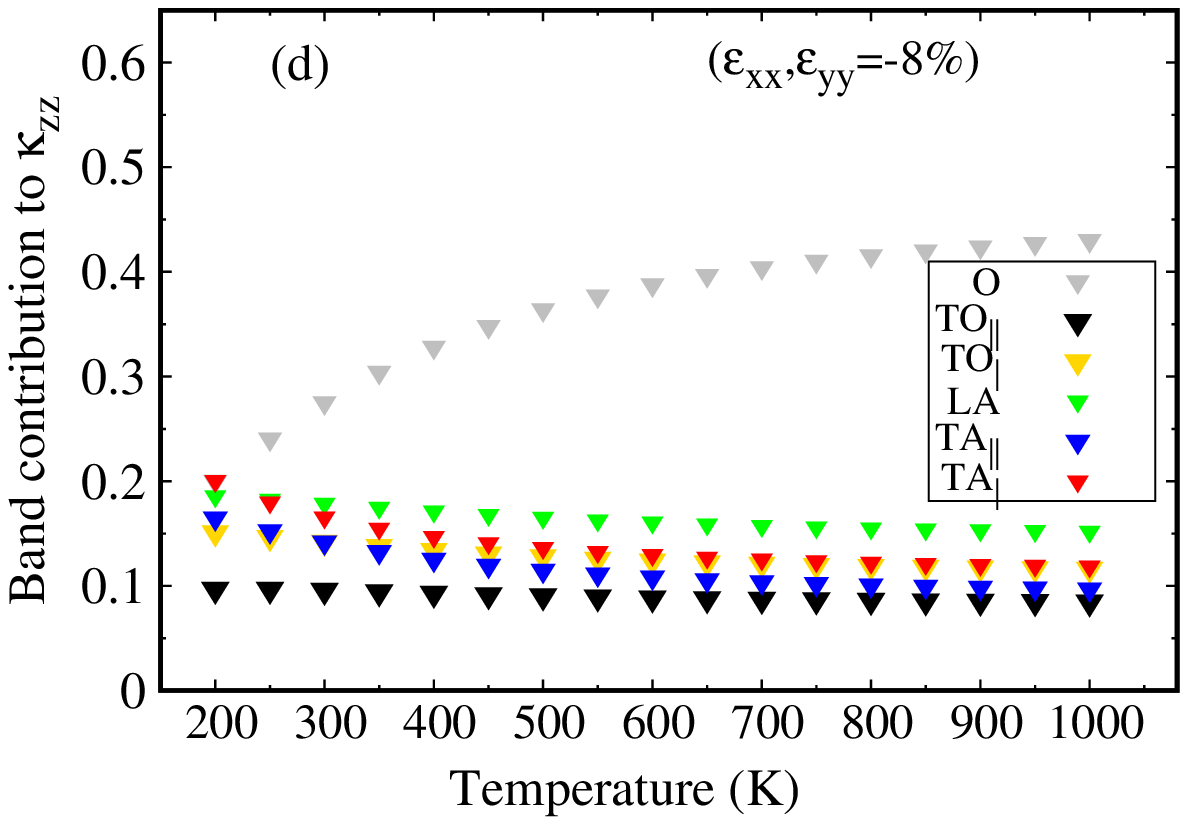}\par
    \includegraphics[height=6 cm,width=1.1\linewidth]{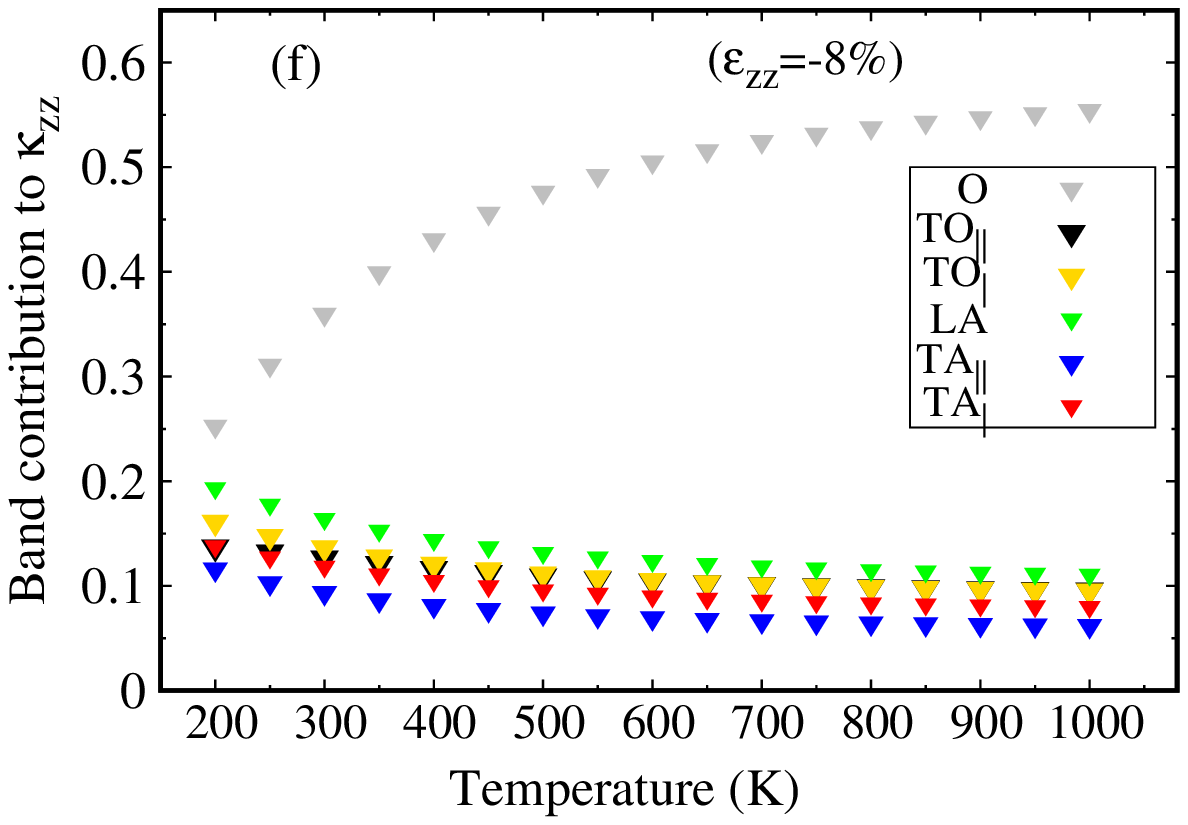}\par 
\end{multicols}
\caption{Normalized contribution of each phonon branch to the total thermal conductivity as a function of temperature. The top three plots show the in-plane thermal conductivity $\kappa_{xx}$ (a, c, and e) while the bottom three show the out-of-plane normalized thermal conductivity $\kappa_{zz}$ (b, d and f). From left to right, we show the contributions for 8\% strain, in all three directions (a and b), in-plane (c and d) and out of plane (e and f) strain respectively.}
  \label{mode-contribution}
\end{figure*}

Finally, we show in Fig.\ref{mode-contribution}  the contribution of different phonon branches to the lattice thermal conductivity of BC$_{2}$N for different pressures. It can clearly be seen that the contribution of "other" optical phonons increases with temperature. This increase is largest for cross-plane compressions (e,f). The relative contributions do not seem to have changed much compared to the zero pressure case of Fig.\ref{mode-contributions}. 

\section{Conclusion}
In this work, thermal properties  of hexagonal BC$_{2}$N, the most energetically stable B-C-N structure reported so far, were calculated. It was found that, due to the strong bond stiffness and light mass of atoms B-C-N, optical phonon frequencies, group velocities and finally the thermal conductivity are quite high and comparable to C-diamond and c-BN. Some optical phonons have group velocities similar to acoustic ones, which makes them have a large contribution to the thermal conductivity. Under 8\% isotropic compressive strain, the lattice thermal conductivity, $\kappa$, of this material increases to 4200 W/m-K, about a factor of 2 larger than the unstrained value at room temperature. This is explained by the shift of phonon modes to higher frequencies resulting in higher  group velocities and a reduction of three-phonon scattering rates due to a reduced scattering phase space. While it was shown that hexagonal BC$_{2}$N was harder than c-BN, the large values of $\kappa$ predicted by the first-principles theory presented here would be interesting to measure under high pressure. Finally, the larger thermal expansion coefficient, compared to diamond and c-BN makes this material attractive as a heat sink with smaller thermal strains when interfaced with Si or polycrystalline 3C-SiC which have almost no CTE mismatch with BC$_2$N. Due to its large bandgap, Si-matching CTE and high thermal conductivity, this material can find applications in high-power electronic devices.

\section{ACKNOWLEDGMENTS}
We would like to acknowledge Prof. Andrea Dal Corso
for his invaluable advice about using Thermo pw, and
the support of the School of Engineering and Applied
Sciences of the University of Virginia for allocating computational resources on the Rivanna supercomputer.


\bibliographystyle{model1-num-names}
\bibliography{myreferences}

\begin{thebibliography}{84}
\expandafter\ifx\csname natexlab\endcsname\relax\def\natexlab#1{#1}\fi
\providecommand{\url}[1]{\texttt{#1}}
\providecommand{\href}[2]{#2}
\providecommand{\path}[1]{#1}
\providecommand{\DOIprefix}{doi:}
\providecommand{\ArXivprefix}{arXiv:}
\providecommand{\URLprefix}{URL: }
\providecommand{\Pubmedprefix}{pmid:}
\providecommand{\doi}[1]{\href{http://dx.doi.org/#1}{\path{#1}}}
\providecommand{\Pubmed}[1]{\href{pmid:#1}{\path{#1}}}
\providecommand{\bibinfo}[2]{#2}
\ifx\xfnm\relax \def\xfnm[#1]{\unskip,\space#1}\fi
\bibitem[{Teter(1998)}]{teter1998computational}
\bibinfo{author}{D.~M. Teter},
\newblock \bibinfo{title}{Computational alchemy: the search for new superhard
  materials},
\newblock \bibinfo{journal}{Mrs Bulletin} \bibinfo{volume}{23}
  (\bibinfo{year}{1998}) \bibinfo{pages}{22--27}.
\bibitem[{Onn et~al.(1992)Onn, Witek, Qiu, Anthony, and
  Banholzer}]{onn1992some}
\bibinfo{author}{D.~Onn}, \bibinfo{author}{A.~Witek}, \bibinfo{author}{Y.~Qiu},
  \bibinfo{author}{T.~Anthony}, \bibinfo{author}{W.~Banholzer},
\newblock \bibinfo{title}{Some aspects of the thermal conductivity of
  isotopically enriched diamond single crystals},
\newblock \bibinfo{journal}{Physical review letters} \bibinfo{volume}{68}
  (\bibinfo{year}{1992}) \bibinfo{pages}{2806}.
\bibitem[{Olson et~al.(1993)Olson, Pohl, Vandersande, Zoltan, Anthony, and
  Banholzer}]{olson1993thermal}
\bibinfo{author}{J.~Olson}, \bibinfo{author}{R.~Pohl},
  \bibinfo{author}{J.~Vandersande}, \bibinfo{author}{A.~Zoltan},
  \bibinfo{author}{T.~Anthony}, \bibinfo{author}{W.~Banholzer},
\newblock \bibinfo{title}{Thermal conductivity of diamond between 170 and 1200
  k and the isotope effect},
\newblock \bibinfo{journal}{Physical Review B} \bibinfo{volume}{47}
  (\bibinfo{year}{1993}) \bibinfo{pages}{14850}.
\bibitem[{Wei et~al.(1993)Wei, Kuo, Thomas, Anthony, and
  Banholzer}]{wei1993thermal}
\bibinfo{author}{L.~Wei}, \bibinfo{author}{P.~Kuo},
  \bibinfo{author}{R.~Thomas}, \bibinfo{author}{T.~Anthony},
  \bibinfo{author}{W.~Banholzer},
\newblock \bibinfo{title}{Thermal conductivity of isotopically modified single
  crystal diamond},
\newblock \bibinfo{journal}{Physical Review Letters} \bibinfo{volume}{70}
  (\bibinfo{year}{1993}) \bibinfo{pages}{3764}.
\bibitem[{Chung et~al.(2007)Chung, Weinberger, Levine, Kavner, Yang, Tolbert,
  and Kaner}]{chung2007synthesis}
\bibinfo{author}{H.-Y. Chung}, \bibinfo{author}{M.~B. Weinberger},
  \bibinfo{author}{J.~B. Levine}, \bibinfo{author}{A.~Kavner},
  \bibinfo{author}{J.-M. Yang}, \bibinfo{author}{S.~H. Tolbert},
  \bibinfo{author}{R.~B. Kaner},
\newblock \bibinfo{title}{Synthesis of ultra-incompressible superhard rhenium
  diboride at ambient pressure},
\newblock \bibinfo{journal}{Science} \bibinfo{volume}{316}
  (\bibinfo{year}{2007}) \bibinfo{pages}{436--439}.
\bibitem[{Kaner(2005)}]{kaner2005rb}
\bibinfo{author}{R.~Kaner},
\newblock \bibinfo{title}{Rb kaner, jj gilman, and sh tolbert, science 308,
  1268 (2005).},
\newblock \bibinfo{journal}{Science} \bibinfo{volume}{308}
  (\bibinfo{year}{2005}) \bibinfo{pages}{1268}.
\bibitem[{Wentorf~Jr(1957)}]{wentorf1957cubic}
\bibinfo{author}{R.~Wentorf~Jr},
\newblock \bibinfo{title}{Cubic form of boron nitride},
\newblock \bibinfo{journal}{The Journal of Chemical Physics}
  \bibinfo{volume}{26} (\bibinfo{year}{1957}) \bibinfo{pages}{956--956}.
\bibitem[{Lindsay et~al.(2013)Lindsay, Broido, and Reinecke}]{lindsay2013first}
\bibinfo{author}{L.~Lindsay}, \bibinfo{author}{D.~Broido},
  \bibinfo{author}{T.~Reinecke},
\newblock \bibinfo{title}{First-principles determination of ultrahigh thermal
  conductivity of boron arsenide: A competitor for diamond?},
\newblock \bibinfo{journal}{Physical review letters} \bibinfo{volume}{111}
  (\bibinfo{year}{2013}) \bibinfo{pages}{025901}.
\bibitem[{Mukhopadhyay and Stewart(2014)}]{mukhopadhyay2014polar}
\bibinfo{author}{S.~Mukhopadhyay}, \bibinfo{author}{D.~A. Stewart},
\newblock \bibinfo{title}{Polar effects on the thermal conductivity of cubic
  boron nitride under pressure},
\newblock \bibinfo{journal}{Physical review letters} \bibinfo{volume}{113}
  (\bibinfo{year}{2014}) \bibinfo{pages}{025901}.
\bibitem[{Slack(1973)}]{slack1973nonmetallic}
\bibinfo{author}{G.~A. Slack},
\newblock \bibinfo{title}{Nonmetallic crystals with high thermal conductivity},
\newblock \bibinfo{journal}{Journal of Physics and Chemistry of Solids}
  \bibinfo{volume}{34} (\bibinfo{year}{1973}) \bibinfo{pages}{321--335}.
\bibitem[{Ward et~al.(2009)Ward, Broido, Stewart, and Deinzer}]{ward2009ab}
\bibinfo{author}{A.~Ward}, \bibinfo{author}{D.~Broido}, \bibinfo{author}{D.~A.
  Stewart}, \bibinfo{author}{G.~Deinzer},
\newblock \bibinfo{title}{Ab initio theory of the lattice thermal conductivity
  in diamond},
\newblock \bibinfo{journal}{Physical Review B} \bibinfo{volume}{80}
  (\bibinfo{year}{2009}) \bibinfo{pages}{125203}.
\bibitem[{Zhao et~al.(2002)Zhao, He, Daemen, Shen, Schwarz, Zhu, Bish, Huang,
  Zhang, Shen et~al.}]{zhao2002superhard}
\bibinfo{author}{Y.~Zhao}, \bibinfo{author}{D.~He},
  \bibinfo{author}{L.~Daemen}, \bibinfo{author}{T.~Shen},
  \bibinfo{author}{R.~Schwarz}, \bibinfo{author}{Y.~Zhu},
  \bibinfo{author}{D.~Bish}, \bibinfo{author}{J.~Huang},
  \bibinfo{author}{J.~Zhang}, \bibinfo{author}{G.~Shen}, et~al.,
\newblock \bibinfo{title}{Superhard b--c--n materials synthesized in
  nanostructured bulks},
\newblock \bibinfo{journal}{Journal of materials research} \bibinfo{volume}{17}
  (\bibinfo{year}{2002}) \bibinfo{pages}{3139--3145}.
\bibitem[{Solozhenko et~al.(2001)Solozhenko, Andrault, Fiquet, Mezouar, and
  Rubie}]{solozhenko2001synthesis}
\bibinfo{author}{V.~L. Solozhenko}, \bibinfo{author}{D.~Andrault},
  \bibinfo{author}{G.~Fiquet}, \bibinfo{author}{M.~Mezouar},
  \bibinfo{author}{D.~C. Rubie},
\newblock \bibinfo{title}{Synthesis of superhard cubic bc 2 n},
\newblock \bibinfo{journal}{Applied Physics Letters} \bibinfo{volume}{78}
  (\bibinfo{year}{2001}) \bibinfo{pages}{1385--1387}.
\bibitem[{Luo et~al.(2007)Luo, Guo, Xu, Wu, Hu, Liu, He, Yu, Tian, and
  Wang}]{luo2007body}
\bibinfo{author}{X.~Luo}, \bibinfo{author}{X.~Guo}, \bibinfo{author}{B.~Xu},
  \bibinfo{author}{Q.~Wu}, \bibinfo{author}{Q.~Hu}, \bibinfo{author}{Z.~Liu},
  \bibinfo{author}{J.~He}, \bibinfo{author}{D.~Yu}, \bibinfo{author}{Y.~Tian},
  \bibinfo{author}{H.-T. Wang},
\newblock \bibinfo{title}{Body-centered superhard b c 2 n phases from first
  principles},
\newblock \bibinfo{journal}{Physical Review B} \bibinfo{volume}{76}
  (\bibinfo{year}{2007}) \bibinfo{pages}{094103}.
\bibitem[{Solozhenko et~al.(2001)Solozhenko, Dub, and
  Novikov}]{solozhenko2001mechanical}
\bibinfo{author}{V.~L. Solozhenko}, \bibinfo{author}{S.~N. Dub},
  \bibinfo{author}{N.~V. Novikov},
\newblock \bibinfo{title}{Mechanical properties of cubic bc2n, a new superhard
  phase},
\newblock \bibinfo{journal}{Diamond and Related Materials} \bibinfo{volume}{10}
  (\bibinfo{year}{2001}) \bibinfo{pages}{2228--2231}.
\bibitem[{Chakraborty et~al.(2018)Chakraborty, Xiong, Cao, and
  Wang}]{chakraborty2018lattice}
\bibinfo{author}{P.~Chakraborty}, \bibinfo{author}{G.~Xiong},
  \bibinfo{author}{L.~Cao}, \bibinfo{author}{Y.~Wang},
\newblock \bibinfo{title}{Lattice thermal transport in superhard hexagonal
  diamond and wurtzite boron nitride: A comparative study with cubic diamond
  and cubic boron nitride},
\newblock \bibinfo{journal}{Carbon} \bibinfo{volume}{139}
  (\bibinfo{year}{2018}) \bibinfo{pages}{85--93}.
\bibitem[{Luo et~al.(2006)Luo, Zhang, Guo, Zhang, He, Yu, Liu, and
  Tian}]{luo2006synthesis}
\bibinfo{author}{X.~Luo}, \bibinfo{author}{J.~Zhang}, \bibinfo{author}{X.~Guo},
  \bibinfo{author}{G.~Zhang}, \bibinfo{author}{J.~He}, \bibinfo{author}{D.~Yu},
  \bibinfo{author}{Z.~Liu}, \bibinfo{author}{Y.~Tian},
\newblock \bibinfo{title}{Synthesis of b--c--n nanocrystalline particle by
  mechanical alloying and spark plasma sintering},
\newblock \bibinfo{journal}{Journal of materials science} \bibinfo{volume}{41}
  (\bibinfo{year}{2006}) \bibinfo{pages}{8352--8355}.
\bibitem[{Huang et~al.(2004)Huang, Cao, Xiang, Lv, and
  Zhu}]{huang2004synthesis}
\bibinfo{author}{F.~L. Huang}, \bibinfo{author}{C.~B. Cao},
  \bibinfo{author}{X.~Xiang}, \bibinfo{author}{R.~T. Lv},
  \bibinfo{author}{H.~S. Zhu},
\newblock \bibinfo{title}{Synthesis of hexagonal boron carbonitride phase by
  solvothermal method},
\newblock \bibinfo{journal}{Diamond and related materials} \bibinfo{volume}{13}
  (\bibinfo{year}{2004}) \bibinfo{pages}{1757--1760}.
\bibitem[{Kaner et~al.(1987)Kaner, Kouvetakis, Warble, Sattler, and
  Bartlett}]{kaner1987boron}
\bibinfo{author}{R.~Kaner}, \bibinfo{author}{J.~Kouvetakis},
  \bibinfo{author}{C.~Warble}, \bibinfo{author}{M.~Sattler},
  \bibinfo{author}{N.~Bartlett},
\newblock \bibinfo{title}{Boron-carbon-nitrogen materials of graphite-like
  structure},
\newblock \bibinfo{journal}{Materials research bulletin} \bibinfo{volume}{22}
  (\bibinfo{year}{1987}) \bibinfo{pages}{399--404}.
\bibitem[{Yamada(1998)}]{yamada1998shock}
\bibinfo{author}{K.~Yamada},
\newblock \bibinfo{title}{Shock synthesis of a graphitic boron-carbon-nitrogen
  system},
\newblock \bibinfo{journal}{Journal of the American Ceramic Society}
  \bibinfo{volume}{81} (\bibinfo{year}{1998}) \bibinfo{pages}{1941--1944}.
\bibitem[{Knittle et~al.(1995)Knittle, Kaner, Jeanloz, and
  Cohen}]{knittle1995high}
\bibinfo{author}{E.~Knittle}, \bibinfo{author}{R.~Kaner},
  \bibinfo{author}{R.~Jeanloz}, \bibinfo{author}{M.~Cohen},
\newblock \bibinfo{title}{High-pressure synthesis, characterization, and
  equation of state of cubic c-bn solid solutions},
\newblock \bibinfo{journal}{Physical Review B} \bibinfo{volume}{51}
  (\bibinfo{year}{1995}) \bibinfo{pages}{12149}.
\bibitem[{Nakano et~al.(1994)Nakano, Akaishi, Sasaki, and
  Yamaoka}]{nakano1994segregative}
\bibinfo{author}{S.~Nakano}, \bibinfo{author}{M.~Akaishi},
  \bibinfo{author}{T.~Sasaki}, \bibinfo{author}{S.~Yamaoka},
\newblock \bibinfo{title}{Segregative crystallization of several diamond-like
  phases from the graphitic bc2n without an additive at 7.7 gpa},
\newblock \bibinfo{journal}{Chemistry of materials} \bibinfo{volume}{6}
  (\bibinfo{year}{1994}) \bibinfo{pages}{2246--2251}.
\bibitem[{Komatsu et~al.(1996)Komatsu, Nomura, Kakudate, and
  Fujiwara}]{komatsu1996synthesis}
\bibinfo{author}{T.~Komatsu}, \bibinfo{author}{M.~Nomura},
  \bibinfo{author}{Y.~Kakudate}, \bibinfo{author}{S.~Fujiwara},
\newblock \bibinfo{title}{Synthesis and characterization of a shock-synthesized
  cubic b--c--n solid solution of composition bc2. 5n},
\newblock \bibinfo{journal}{Journal of Materials Chemistry} \bibinfo{volume}{6}
  (\bibinfo{year}{1996}) \bibinfo{pages}{1799--1803}.
\bibitem[{Yao et~al.(1998)Yao, Chen, Liu, Ding, and Su}]{yao1998amorphous}
\bibinfo{author}{B.~Yao}, \bibinfo{author}{W.~Chen}, \bibinfo{author}{L.~Liu},
  \bibinfo{author}{B.~Ding}, \bibinfo{author}{W.~Su},
\newblock \bibinfo{title}{Amorphous b--c--n semiconductor},
\newblock \bibinfo{journal}{Journal of applied physics} \bibinfo{volume}{84}
  (\bibinfo{year}{1998}) \bibinfo{pages}{1412--1415}.
\bibitem[{Tkachev et~al.(2003)Tkachev, Solozhenko, Zinin, Manghnani, and
  Ming}]{tkachev2003elastic}
\bibinfo{author}{S.~Tkachev}, \bibinfo{author}{V.~Solozhenko},
  \bibinfo{author}{P.~Zinin}, \bibinfo{author}{M.~Manghnani},
  \bibinfo{author}{L.~Ming},
\newblock \bibinfo{title}{Elastic moduli of the superhard cubic bc 2 n phase by
  brillouin scattering},
\newblock \bibinfo{journal}{Physical Review B} \bibinfo{volume}{68}
  (\bibinfo{year}{2003}) \bibinfo{pages}{052104}.
\bibitem[{Wu et~al.(2006)Wu, Liu, Hu, Li, He, Yu, Li, and Tian}]{wu2006thermal}
\bibinfo{author}{Q.~Wu}, \bibinfo{author}{Z.~Liu}, \bibinfo{author}{Q.~Hu},
  \bibinfo{author}{H.~Li}, \bibinfo{author}{J.~He}, \bibinfo{author}{D.~Yu},
  \bibinfo{author}{D.~Li}, \bibinfo{author}{Y.~Tian},
\newblock \bibinfo{title}{The thermal expansion of a highly crystalline
  hexagonal bc2n compound synthesized under high temperature and pressure},
\newblock \bibinfo{journal}{Journal of Physics: Condensed Matter}
  \bibinfo{volume}{18} (\bibinfo{year}{2006}) \bibinfo{pages}{9519}.
\bibitem[{Kouvetakis et~al.(1989)Kouvetakis, Sasaki, Shen, Hagiwara, Lerner,
  Krishnan, and Bartlett}]{kouvetakis1989novel}
\bibinfo{author}{J.~Kouvetakis}, \bibinfo{author}{T.~Sasaki},
  \bibinfo{author}{C.~Shen}, \bibinfo{author}{R.~Hagiwara},
  \bibinfo{author}{M.~Lerner}, \bibinfo{author}{K.~Krishnan},
  \bibinfo{author}{N.~Bartlett},
\newblock \bibinfo{title}{Novel aspects of graphite intercalation by fluorine
  and fluorides and new b/c, c/n and b/c/n materials based on the graphite
  network},
\newblock \bibinfo{journal}{Synthetic metals} \bibinfo{volume}{34}
  (\bibinfo{year}{1989}) \bibinfo{pages}{1--7}.
\bibitem[{Solozhenko et~al.(1997)Solozhenko, Turkevich, and
  Sato}]{solozhenko1997phase}
\bibinfo{author}{V.~L. Solozhenko}, \bibinfo{author}{V.~Z. Turkevich},
  \bibinfo{author}{T.~Sato},
\newblock \bibinfo{title}{Phase stability of graphitelike bc4n up to 2100 k and
  7 gpa},
\newblock \bibinfo{journal}{Journal of the American Ceramic Society}
  \bibinfo{volume}{80} (\bibinfo{year}{1997}) \bibinfo{pages}{3229--3232}.
\bibitem[{Hub{\'a}{\v{c}}ek and Sato(1995)}]{hubavcek1995preparation}
\bibinfo{author}{M.~Hub{\'a}{\v{c}}ek}, \bibinfo{author}{T.~Sato},
\newblock \bibinfo{title}{Preparation and properties of a compound in the bcn
  system},
\newblock \bibinfo{journal}{Journal of Solid State Chemistry}
  \bibinfo{volume}{114} (\bibinfo{year}{1995}) \bibinfo{pages}{258--264}.
\bibitem[{He et~al.(2001)He, Tian, Yu, Wang, Liu, Guo, Li, Jia, Chen, Zou
  et~al.}]{he2001orthorhombic}
\bibinfo{author}{J.~He}, \bibinfo{author}{Y.~Tian}, \bibinfo{author}{D.~Yu},
  \bibinfo{author}{T.~Wang}, \bibinfo{author}{S.~Liu},
  \bibinfo{author}{L.~Guo}, \bibinfo{author}{D.~Li}, \bibinfo{author}{X.~Jia},
  \bibinfo{author}{L.~Chen}, \bibinfo{author}{G.~Zou}, et~al.,
\newblock \bibinfo{title}{Orthorhombic b2cn crystal synthesized by high
  pressure and temperature},
\newblock \bibinfo{journal}{Chemical physics letters} \bibinfo{volume}{340}
  (\bibinfo{year}{2001}) \bibinfo{pages}{431--436}.
\bibitem[{Kim et~al.(2007)Kim, Pang, Utsumi, Solozhenko, and
  Zhao}]{kim2007cubic}
\bibinfo{author}{E.~Kim}, \bibinfo{author}{T.~Pang},
  \bibinfo{author}{W.~Utsumi}, \bibinfo{author}{V.~L. Solozhenko},
  \bibinfo{author}{Y.~Zhao},
\newblock \bibinfo{title}{Cubic phases of b c 2 n: A first-principles study},
\newblock \bibinfo{journal}{Physical Review B} \bibinfo{volume}{75}
  (\bibinfo{year}{2007}) \bibinfo{pages}{184115}.
\bibitem[{Mattesini and Matar(2001)}]{mattesini2001search}
\bibinfo{author}{M.~Mattesini}, \bibinfo{author}{S.~Matar},
\newblock \bibinfo{title}{Search for ultra-hard materials: theoretical
  characterisation of novel orthorhombic bc2n crystals},
\newblock \bibinfo{journal}{International Journal of Inorganic Materials}
  \bibinfo{volume}{3} (\bibinfo{year}{2001}) \bibinfo{pages}{943--957}.
\bibitem[{Zhang et~al.(1999)Zhang, Chan, Cheung, and Lee}]{zhang1999energetics}
\bibinfo{author}{R.~Zhang}, \bibinfo{author}{K.~Chan},
  \bibinfo{author}{H.~Cheung}, \bibinfo{author}{S.~Lee},
\newblock \bibinfo{title}{Energetics of segregation in $\beta$-c 2 bn},
\newblock \bibinfo{journal}{Applied physics letters} \bibinfo{volume}{75}
  (\bibinfo{year}{1999}) \bibinfo{pages}{2259--2261}.
\bibitem[{Pan et~al.(2005)Pan, Sun, and Chen}]{pan2005ab}
\bibinfo{author}{Z.~Pan}, \bibinfo{author}{H.~Sun}, \bibinfo{author}{C.~Chen},
\newblock \bibinfo{title}{Ab initio pseudopotential studies of cubic bc2n under
  high pressure},
\newblock \bibinfo{journal}{Journal of Physics: Condensed Matter}
  \bibinfo{volume}{17} (\bibinfo{year}{2005}) \bibinfo{pages}{3211}.
\bibitem[{Mattesini and Matar(2001)}]{mattesini2001first}
\bibinfo{author}{M.~Mattesini}, \bibinfo{author}{S.~F. Matar},
\newblock \bibinfo{title}{First-principles characterisation of new ternary
  heterodiamond bc2n phases},
\newblock \bibinfo{journal}{Computational materials science}
  \bibinfo{volume}{20} (\bibinfo{year}{2001}) \bibinfo{pages}{107--119}.
\bibitem[{Widany et~al.(1998)Widany, Verwoerd, and
  Frauenheim}]{widany1998density}
\bibinfo{author}{J.~Widany}, \bibinfo{author}{W.~Verwoerd},
  \bibinfo{author}{T.~Frauenheim},
\newblock \bibinfo{title}{Density-functional based tight-binding calculations
  on zinc--blende type bc2n-crystals},
\newblock \bibinfo{journal}{Diamond and related materials} \bibinfo{volume}{7}
  (\bibinfo{year}{1998}) \bibinfo{pages}{1633--1638}.
\bibitem[{Lambrecht and Segall(1989)}]{lambrecht1989electronic}
\bibinfo{author}{W.~R. Lambrecht}, \bibinfo{author}{B.~Segall},
\newblock \bibinfo{title}{Electronic structure of (diamond c)/(sphalerite
  bn)(110) interfaces and superlattices},
\newblock \bibinfo{journal}{Physical Review B} \bibinfo{volume}{40}
  (\bibinfo{year}{1989}) \bibinfo{pages}{9909}.
\bibitem[{Tateyama et~al.(1997)Tateyama, Ogitsu, Kusakabe, Tsuneyuki, and
  Itoh}]{tateyama1997proposed}
\bibinfo{author}{Y.~Tateyama}, \bibinfo{author}{T.~Ogitsu},
  \bibinfo{author}{K.~Kusakabe}, \bibinfo{author}{S.~Tsuneyuki},
  \bibinfo{author}{S.~Itoh},
\newblock \bibinfo{title}{Proposed synthesis path for heterodiamond bc 2 n},
\newblock \bibinfo{journal}{Physical Review B} \bibinfo{volume}{55}
  (\bibinfo{year}{1997}) \bibinfo{pages}{R10161}.
\bibitem[{Sun et~al.(2001)Sun, Jhi, Roundy, Cohen, and
  Louie}]{sun2001structural}
\bibinfo{author}{H.~Sun}, \bibinfo{author}{S.-H. Jhi},
  \bibinfo{author}{D.~Roundy}, \bibinfo{author}{M.~L. Cohen},
  \bibinfo{author}{S.~G. Louie},
\newblock \bibinfo{title}{Structural forms of cubic bc 2 n},
\newblock \bibinfo{journal}{Physical Review B} \bibinfo{volume}{64}
  (\bibinfo{year}{2001}) \bibinfo{pages}{094108}.
\bibitem[{Sun et~al.(2006)Sun, Zhou, Fan, Chen, Wang, Guo, He, and
  Tian}]{sun2006first}
\bibinfo{author}{J.~Sun}, \bibinfo{author}{X.-F. Zhou}, \bibinfo{author}{Y.-X.
  Fan}, \bibinfo{author}{J.~Chen}, \bibinfo{author}{H.-T. Wang},
  \bibinfo{author}{X.~Guo}, \bibinfo{author}{J.~He}, \bibinfo{author}{Y.~Tian},
\newblock \bibinfo{title}{First-principles study of electronic structure and
  optical properties of heterodiamond bc 2 n},
\newblock \bibinfo{journal}{Physical Review B} \bibinfo{volume}{73}
  (\bibinfo{year}{2006}) \bibinfo{pages}{045108}.
\bibitem[{Zhang et~al.(2004)Zhang, Sun, and Chen}]{zhang2004superhard}
\bibinfo{author}{Y.~Zhang}, \bibinfo{author}{H.~Sun},
  \bibinfo{author}{C.~Chen},
\newblock \bibinfo{title}{Superhard cubic b c 2 n compared to diamond},
\newblock \bibinfo{journal}{Physical review letters} \bibinfo{volume}{93}
  (\bibinfo{year}{2004}) \bibinfo{pages}{195504}.
\bibitem[{Sun et~al.(2006)Sun, Zhou, Qian, Chen, Fan, Wang, Guo, He, Liu, and
  Tian}]{sun2006chalcopyrite}
\bibinfo{author}{J.~Sun}, \bibinfo{author}{X.-F. Zhou}, \bibinfo{author}{G.-R.
  Qian}, \bibinfo{author}{J.~Chen}, \bibinfo{author}{Y.-X. Fan},
  \bibinfo{author}{H.-T. Wang}, \bibinfo{author}{X.~Guo},
  \bibinfo{author}{J.~He}, \bibinfo{author}{Z.~Liu}, \bibinfo{author}{Y.~Tian},
\newblock \bibinfo{title}{Chalcopyrite polymorph for superhard bc 2 n},
\newblock \bibinfo{journal}{Applied physics letters} \bibinfo{volume}{89}
  (\bibinfo{year}{2006}) \bibinfo{pages}{151911}.
\bibitem[{Zhou et~al.(2007)Zhou, Sun, Fan, Chen, Wang, Guo, He, and
  Tian}]{zhou2007most}
\bibinfo{author}{X.-F. Zhou}, \bibinfo{author}{J.~Sun}, \bibinfo{author}{Y.-X.
  Fan}, \bibinfo{author}{J.~Chen}, \bibinfo{author}{H.-T. Wang},
  \bibinfo{author}{X.~Guo}, \bibinfo{author}{J.~He}, \bibinfo{author}{Y.~Tian},
\newblock \bibinfo{title}{Most likely phase of superhard bc 2 n by ab initio
  calculations},
\newblock \bibinfo{journal}{Physical Review B} \bibinfo{volume}{76}
  (\bibinfo{year}{2007}) \bibinfo{pages}{100101}.
\bibitem[{Luo et~al.(2007)Luo, Guo, Liu, He, Yu, Xu, Tian, and
  Wang}]{luo2007first}
\bibinfo{author}{X.~Luo}, \bibinfo{author}{X.~Guo}, \bibinfo{author}{Z.~Liu},
  \bibinfo{author}{J.~He}, \bibinfo{author}{D.~Yu}, \bibinfo{author}{B.~Xu},
  \bibinfo{author}{Y.~Tian}, \bibinfo{author}{H.-T. Wang},
\newblock \bibinfo{title}{First-principles study of wurtzite b c 2 n},
\newblock \bibinfo{journal}{Physical Review B} \bibinfo{volume}{76}
  (\bibinfo{year}{2007}) \bibinfo{pages}{092107}.
\bibitem[{Chen et~al.(2007)Chen, Gong, and Wei}]{chen2007superhard}
\bibinfo{author}{S.~Chen}, \bibinfo{author}{X.~Gong}, \bibinfo{author}{S.-H.
  Wei},
\newblock \bibinfo{title}{Superhard pseudocubic bc 2 n superlattices},
\newblock \bibinfo{journal}{Physical review letters} \bibinfo{volume}{98}
  (\bibinfo{year}{2007}) \bibinfo{pages}{015502}.
\bibitem[{Chen et~al.(2008)Chen, Gong, and Wei}]{chen2008crystal}
\bibinfo{author}{S.~Chen}, \bibinfo{author}{X.~Gong}, \bibinfo{author}{S.-H.
  Wei},
\newblock \bibinfo{title}{Crystal structures and mechanical properties of
  superhard bc$_2$n and bc$_4$n alloys: First-principles calculations},
\newblock \bibinfo{journal}{Physical Review B} \bibinfo{volume}{77}
  (\bibinfo{year}{2008}) \bibinfo{pages}{014113}.
\bibitem[{Chen et~al.(2007)Chen, Gong, and Wei}]{chen2007chen}
\bibinfo{author}{S.~Chen}, \bibinfo{author}{X.~Gong}, \bibinfo{author}{S.-H.
  Wei},
\newblock \bibinfo{title}{Chen, gong, and wei reply},
\newblock \bibinfo{journal}{Physical Review Letters} \bibinfo{volume}{99}
  (\bibinfo{year}{2007}) \bibinfo{pages}{159602}.
\bibitem[{Zhou et~al.(2009)Zhou, Sun, Qian, Guo, Liu, Tian, and
  Wang}]{zhou2009tetragonal}
\bibinfo{author}{X.-F. Zhou}, \bibinfo{author}{J.~Sun}, \bibinfo{author}{Q.-R.
  Qian}, \bibinfo{author}{X.~Guo}, \bibinfo{author}{Z.~Liu},
  \bibinfo{author}{Y.~Tian}, \bibinfo{author}{H.-T. Wang},
\newblock \bibinfo{title}{A tetragonal phase of superhard bc 2 n},
\newblock \bibinfo{journal}{Journal of applied physics} \bibinfo{volume}{105}
  (\bibinfo{year}{2009}) \bibinfo{pages}{093521}.
\bibitem[{Li et~al.(2009)Li, Wang, Oganov, Cui, Ma, and
  Zou}]{li2009rhombohedral}
\bibinfo{author}{Q.~Li}, \bibinfo{author}{M.~Wang}, \bibinfo{author}{A.~R.
  Oganov}, \bibinfo{author}{T.~Cui}, \bibinfo{author}{Y.~Ma},
  \bibinfo{author}{G.~Zou},
\newblock \bibinfo{title}{Rhombohedral superhard structure of bc 2 n},
\newblock \bibinfo{journal}{Journal of Applied Physics} \bibinfo{volume}{105}
  (\bibinfo{year}{2009}) \bibinfo{pages}{053514}.
\bibitem[{Oganov and Glass(2006)}]{oganov2006crystal}
\bibinfo{author}{A.~R. Oganov}, \bibinfo{author}{C.~W. Glass},
\newblock \bibinfo{title}{Crystal structure prediction using ab initio
  evolutionary techniques: Principles and applications},
\newblock \bibinfo{journal}{The Journal of chemical physics}
  \bibinfo{volume}{124} (\bibinfo{year}{2006}) \bibinfo{pages}{244704}.
\bibitem[{Glass et~al.(2006)Glass, Oganov, and Hansen}]{glass2006uspex}
\bibinfo{author}{C.~W. Glass}, \bibinfo{author}{A.~R. Oganov},
  \bibinfo{author}{N.~Hansen},
\newblock \bibinfo{title}{Uspex—evolutionary crystal structure prediction},
\newblock \bibinfo{journal}{Computer physics communications}
  \bibinfo{volume}{175} (\bibinfo{year}{2006}) \bibinfo{pages}{713--720}.
\bibitem[{Oganov et~al.(2006)Oganov, Glass, and Ono}]{oganov2006high}
\bibinfo{author}{A.~R. Oganov}, \bibinfo{author}{C.~W. Glass},
  \bibinfo{author}{S.~Ono},
\newblock \bibinfo{title}{High-pressure phases of caco3: crystal structure
  prediction and experiment},
\newblock \bibinfo{journal}{Earth and Planetary Science Letters}
  \bibinfo{volume}{241} (\bibinfo{year}{2006}) \bibinfo{pages}{95--103}.
\bibitem[{Liu et~al.(2018)Liu, Zhao, Yu, Zhang, Lin, and
  Yang}]{liu2018hexagonal}
\bibinfo{author}{L.~Liu}, \bibinfo{author}{Z.~Zhao}, \bibinfo{author}{T.~Yu},
  \bibinfo{author}{S.~Zhang}, \bibinfo{author}{J.~Lin},
  \bibinfo{author}{G.~Yang},
\newblock \bibinfo{title}{Hexagonal bc2n with remarkably high hardness},
\newblock \bibinfo{journal}{The Journal of Physical Chemistry C}
  \bibinfo{volume}{122} (\bibinfo{year}{2018}) \bibinfo{pages}{6801--6807}.
\bibitem[{Broido et~al.(2007)Broido, Malorny, Birner, Mingo, and
  Stewart}]{broido2007intrinsic}
\bibinfo{author}{D.~A. Broido}, \bibinfo{author}{M.~Malorny},
  \bibinfo{author}{G.~Birner}, \bibinfo{author}{N.~Mingo},
  \bibinfo{author}{D.~Stewart},
\newblock \bibinfo{title}{Intrinsic lattice thermal conductivity of
  semiconductors from first principles},
\newblock \bibinfo{journal}{Applied Physics Letters} \bibinfo{volume}{91}
  (\bibinfo{year}{2007}) \bibinfo{pages}{231922}.
\bibitem[{Omini and Sparavigna(1996)}]{omini1996beyond}
\bibinfo{author}{M.~Omini}, \bibinfo{author}{A.~Sparavigna},
\newblock \bibinfo{title}{Beyond the isotropic-model approximation in the
  theory of thermal conductivity},
\newblock \bibinfo{journal}{Physical Review B} \bibinfo{volume}{53}
  (\bibinfo{year}{1996}) \bibinfo{pages}{9064}.
\bibitem[{Garg et~al.(2011)Garg, Bonini, Kozinsky, and Marzari}]{garg2011role}
\bibinfo{author}{J.~Garg}, \bibinfo{author}{N.~Bonini},
  \bibinfo{author}{B.~Kozinsky}, \bibinfo{author}{N.~Marzari},
\newblock \bibinfo{title}{Role of disorder and anharmonicity in the thermal
  conductivity of silicon-germanium alloys: A first-principles study},
\newblock \bibinfo{journal}{Physical review letters} \bibinfo{volume}{106}
  (\bibinfo{year}{2011}) \bibinfo{pages}{045901}.
\bibitem[{Esfarjani et~al.(2011)Esfarjani, Chen, and
  Stokes}]{esfarjani2011heat}
\bibinfo{author}{K.~Esfarjani}, \bibinfo{author}{G.~Chen},
  \bibinfo{author}{H.~T. Stokes},
\newblock \bibinfo{title}{Heat transport in silicon from first-principles
  calculations},
\newblock \bibinfo{journal}{Physical Review B} \bibinfo{volume}{84}
  (\bibinfo{year}{2011}) \bibinfo{pages}{085204}.
\bibitem[{Tang and Dong(2010)}]{tang2010lattice}
\bibinfo{author}{X.~Tang}, \bibinfo{author}{J.~Dong},
\newblock \bibinfo{title}{Lattice thermal conductivity of mgo at conditions of
  earth’s interior},
\newblock \bibinfo{journal}{Proceedings of the National Academy of Sciences}
  \bibinfo{volume}{107} (\bibinfo{year}{2010}) \bibinfo{pages}{4539--4543}.
\bibitem[{Huang et~al.(2019)Huang, Fan, Singh, and Zheng}]{huang2019thermal}
\bibinfo{author}{H.~Huang}, \bibinfo{author}{X.~Fan}, \bibinfo{author}{D.~J.
  Singh}, \bibinfo{author}{W.~Zheng},
\newblock \bibinfo{title}{The thermal and thermoelectric transport properties
  of sisb, gesb and snsb monolayers},
\newblock \bibinfo{journal}{Journal of Materials Chemistry C}
  \bibinfo{volume}{7} (\bibinfo{year}{2019}) \bibinfo{pages}{10652--10662}.
\bibitem[{Pandey et~al.(2017)Pandey, Parker, and Lindsay}]{pandey2017ab}
\bibinfo{author}{T.~Pandey}, \bibinfo{author}{D.~S. Parker},
  \bibinfo{author}{L.~Lindsay},
\newblock \bibinfo{title}{Ab initio phonon thermal transport in monolayer inse,
  gase, gas, and alloys},
\newblock \bibinfo{journal}{Nanotechnology} \bibinfo{volume}{28}
  (\bibinfo{year}{2017}) \bibinfo{pages}{455706}.
\bibitem[{Zeraati et~al.(2016)Zeraati, Allaei, Sarsari, Pourfath, and
  Donadio}]{zeraati2016highly}
\bibinfo{author}{M.~Zeraati}, \bibinfo{author}{S.~M.~V. Allaei},
  \bibinfo{author}{I.~A. Sarsari}, \bibinfo{author}{M.~Pourfath},
  \bibinfo{author}{D.~Donadio},
\newblock \bibinfo{title}{Highly anisotropic thermal conductivity of arsenene:
  An ab initio study},
\newblock \bibinfo{journal}{Physical Review B} \bibinfo{volume}{93}
  (\bibinfo{year}{2016}) \bibinfo{pages}{085424}.
\bibitem[{Giannozzi et~al.(2009)Giannozzi, Baroni, Bonini, Calandra, Car,
  Cavazzoni, Ceresoli, Chiarotti, Cococcioni, Dabo
  et~al.}]{giannozzi2009quantum}
\bibinfo{author}{P.~Giannozzi}, \bibinfo{author}{S.~Baroni},
  \bibinfo{author}{N.~Bonini}, \bibinfo{author}{M.~Calandra},
  \bibinfo{author}{R.~Car}, \bibinfo{author}{C.~Cavazzoni},
  \bibinfo{author}{D.~Ceresoli}, \bibinfo{author}{G.~L. Chiarotti},
  \bibinfo{author}{M.~Cococcioni}, \bibinfo{author}{I.~Dabo}, et~al.,
\newblock \bibinfo{title}{Quantum espresso: a modular and open-source software
  project for quantum simulations of materials},
\newblock \bibinfo{journal}{Journal of physics: Condensed matter}
  \bibinfo{volume}{21} (\bibinfo{year}{2009}) \bibinfo{pages}{395502}.
\bibitem[{Togo and Tanaka(2015)}]{phonopy}
\bibinfo{author}{A.~Togo}, \bibinfo{author}{I.~Tanaka},
\newblock \bibinfo{title}{First principles phonon calculations in materials
  science},
\newblock \bibinfo{journal}{Scr. Mater.} \bibinfo{volume}{108}
  (\bibinfo{year}{2015}) \bibinfo{pages}{1--5}.
\bibitem[{Li et~al.(2014)Li, Carrete, Katcho, and Mingo}]{li2014shengbte}
\bibinfo{author}{W.~Li}, \bibinfo{author}{J.~Carrete}, \bibinfo{author}{N.~A.
  Katcho}, \bibinfo{author}{N.~Mingo},
\newblock \bibinfo{title}{Shengbte: A solver of the boltzmann transport
  equation for phonons},
\newblock \bibinfo{journal}{Computer Physics Communications}
  \bibinfo{volume}{185} (\bibinfo{year}{2014}) \bibinfo{pages}{1747--1758}.
\bibitem[{Ziman(1960)}]{ziman2010electrons}
\bibinfo{author}{J.~Ziman}, \bibinfo{title}{Electrons and phonons},
  \bibinfo{year}{1960}.
\bibitem[{Li and Mingo(2014)}]{li2014thermal}
\bibinfo{author}{W.~Li}, \bibinfo{author}{N.~Mingo},
\newblock \bibinfo{title}{Thermal conductivity of fully filled skutterudites:
  Role of the filler},
\newblock \bibinfo{journal}{Physical Review B} \bibinfo{volume}{89}
  (\bibinfo{year}{2014}) \bibinfo{pages}{184304}.
\bibitem[{Li and Mingo(2015)}]{li2015ultralow}
\bibinfo{author}{W.~Li}, \bibinfo{author}{N.~Mingo},
\newblock \bibinfo{title}{Ultralow lattice thermal conductivity of the fully
  filled skutterudite ybfe 4 sb 12 due to the flat avoided-crossing filler
  modes},
\newblock \bibinfo{journal}{Physical Review B} \bibinfo{volume}{91}
  (\bibinfo{year}{2015}) \bibinfo{pages}{144304}.
\bibitem[{Palumbo and Dal~Corso(2017)}]{palumbo2017lattice}
\bibinfo{author}{M.~Palumbo}, \bibinfo{author}{A.~Dal~Corso},
\newblock \bibinfo{title}{Lattice dynamics and thermophysical properties of hcp
  os and ru from the quasi-harmonic approximation},
\newblock \bibinfo{journal}{Journal of Physics: Condensed Matter}
  \bibinfo{volume}{29} (\bibinfo{year}{2017}) \bibinfo{pages}{395401}.
\bibitem[{Mouhat and Coudert(2014)}]{mouhat2014necessary}
\bibinfo{author}{F.~Mouhat}, \bibinfo{author}{F.-X. Coudert},
\newblock \bibinfo{title}{Necessary and sufficient elastic stability conditions
  in various crystal systems},
\newblock \bibinfo{journal}{Physical review B} \bibinfo{volume}{90}
  (\bibinfo{year}{2014}) \bibinfo{pages}{224104}.
\bibitem[{Li et~al.(2019)Li, Shi, Zhu, and Xia}]{li2019elastic}
\bibinfo{author}{S.~Li}, \bibinfo{author}{L.~Shi}, \bibinfo{author}{H.~Zhu},
  \bibinfo{author}{W.~Xia},
\newblock \bibinfo{title}{Elastic and bandgap modulations of hexagonal bc2n
  from first-principles calculations},
\newblock \bibinfo{journal}{physica status solidi (b)}  (\bibinfo{year}{2019})
  \bibinfo{pages}{1900281}.
\bibitem[{Mounet and Marzari(2005)}]{mounet2005first}
\bibinfo{author}{N.~Mounet}, \bibinfo{author}{N.~Marzari},
\newblock \bibinfo{title}{First-principles determination of the structural,
  vibrational and thermodynamic properties of diamond, graphite, and
  derivatives},
\newblock \bibinfo{journal}{Physical Review B} \bibinfo{volume}{71}
  (\bibinfo{year}{2005}) \bibinfo{pages}{205214}.
\bibitem[{Cao et~al.(2019)Cao, Zhao, Zhou, Liu, and Gan}]{cao2019superhard}
\bibinfo{author}{A.-H. Cao}, \bibinfo{author}{W.-J. Zhao},
  \bibinfo{author}{Q.-Y. Zhou}, \bibinfo{author}{S.-L. Liu},
  \bibinfo{author}{L.-H. Gan},
\newblock \bibinfo{title}{A superhard allotrope of carbon: Ibam-c and its bn
  phase},
\newblock \bibinfo{journal}{Chemical Physics Letters} \bibinfo{volume}{714}
  (\bibinfo{year}{2019}) \bibinfo{pages}{119--124}.
\bibitem[{Fan et~al.(2015)Fan, Wei, Chai, Yan, Zhang, Lin, Zhang, Zhang, and
  Zhang}]{fan2015structural}
\bibinfo{author}{Q.~Fan}, \bibinfo{author}{Q.~Wei}, \bibinfo{author}{C.~Chai},
  \bibinfo{author}{H.~Yan}, \bibinfo{author}{M.~Zhang},
  \bibinfo{author}{Z.~Lin}, \bibinfo{author}{Z.~Zhang},
  \bibinfo{author}{J.~Zhang}, \bibinfo{author}{D.~Zhang},
\newblock \bibinfo{title}{Structural, mechanical, and electronic properties of
  p3m1-bcn},
\newblock \bibinfo{journal}{Journal of Physics and Chemistry of Solids}
  \bibinfo{volume}{79} (\bibinfo{year}{2015}) \bibinfo{pages}{89--96}.
\bibitem[{Broido et~al.(2013)Broido, Lindsay, and Reinecke}]{broido2013ab}
\bibinfo{author}{D.~Broido}, \bibinfo{author}{L.~Lindsay},
  \bibinfo{author}{T.~Reinecke},
\newblock \bibinfo{title}{Ab initio study of the unusual thermal transport
  properties of boron arsenide and related materials},
\newblock \bibinfo{journal}{Physical Review B} \bibinfo{volume}{88}
  (\bibinfo{year}{2013}) \bibinfo{pages}{214303}.
\bibitem[{Stoupin and Shvyd’ko(2011)}]{stoupin2011ultraprecise}
\bibinfo{author}{S.~Stoupin}, \bibinfo{author}{Y.~V. Shvyd’ko},
\newblock \bibinfo{title}{Ultraprecise studies of the thermal expansion
  coefficient of diamond using backscattering x-ray diffraction},
\newblock \bibinfo{journal}{Physical Review B} \bibinfo{volume}{83}
  (\bibinfo{year}{2011}) \bibinfo{pages}{104102}.
\bibitem[{Slack and Bartram(1975)}]{slack1975thermal}
\bibinfo{author}{G.~A. Slack}, \bibinfo{author}{S.~Bartram},
\newblock \bibinfo{title}{Thermal expansion of some diamondlike crystals},
\newblock \bibinfo{journal}{Journal of Applied Physics} \bibinfo{volume}{46}
  (\bibinfo{year}{1975}) \bibinfo{pages}{89--98}.
\bibitem[{Okada and Tokumaru(1984)}]{okada1984precise}
\bibinfo{author}{Y.~Okada}, \bibinfo{author}{Y.~Tokumaru},
\newblock \bibinfo{title}{Precise determination of lattice parameter and
  thermal expansion coefficient of silicon between 300 and 1500 k},
\newblock \bibinfo{journal}{Journal of applied physics} \bibinfo{volume}{56}
  (\bibinfo{year}{1984}) \bibinfo{pages}{314--320}.
\bibitem[{Watanabe et~al.(2004)Watanabe, Yamada, and
  Okaji}]{watanabe2004linear}
\bibinfo{author}{H.~Watanabe}, \bibinfo{author}{N.~Yamada},
  \bibinfo{author}{M.~Okaji},
\newblock \bibinfo{title}{Linear thermal expansion coefficient of silicon from
  293 to 1000 k},
\newblock \bibinfo{journal}{International journal of thermophysics}
  \bibinfo{volume}{25} (\bibinfo{year}{2004}) \bibinfo{pages}{221--236}.
\bibitem[{Ward and Broido(2010)}]{ward2010intrinsic}
\bibinfo{author}{A.~Ward}, \bibinfo{author}{D.~Broido},
\newblock \bibinfo{title}{Intrinsic phonon relaxation times from
  first-principles studies of the thermal conductivities of si and ge},
\newblock \bibinfo{journal}{Physical Review B} \bibinfo{volume}{81}
  (\bibinfo{year}{2010}) \bibinfo{pages}{085205}.
\bibitem[{Tian et~al.(2011)Tian, Esfarjani, Shiomi, Henry, and
  Chen}]{tian2011importance}
\bibinfo{author}{Z.~Tian}, \bibinfo{author}{K.~Esfarjani},
  \bibinfo{author}{J.~Shiomi}, \bibinfo{author}{A.~S. Henry},
  \bibinfo{author}{G.~Chen},
\newblock \bibinfo{title}{On the importance of optical phonons to thermal
  conductivity in nanostructures},
\newblock \bibinfo{journal}{Applied Physics Letters} \bibinfo{volume}{99}
  (\bibinfo{year}{2011}) \bibinfo{pages}{053122}.
\bibitem[{Wang et~al.(2004)Wang, Hsu, Pu, Sung, and
  Hwa}]{wang2004determination}
\bibinfo{author}{S.-F. Wang}, \bibinfo{author}{Y.-F. Hsu},
  \bibinfo{author}{J.-C. Pu}, \bibinfo{author}{J.~C. Sung},
  \bibinfo{author}{L.~Hwa},
\newblock \bibinfo{title}{Determination of acoustic wave velocities and elastic
  properties for diamond and other hard materials},
\newblock \bibinfo{journal}{Materials Chemistry and Physics}
  \bibinfo{volume}{85} (\bibinfo{year}{2004}) \bibinfo{pages}{432--437}.
\bibitem[{Broido et~al.(2012)Broido, Lindsay, and Ward}]{broido2012thermal}
\bibinfo{author}{D.~Broido}, \bibinfo{author}{L.~Lindsay},
  \bibinfo{author}{A.~Ward},
\newblock \bibinfo{title}{Thermal conductivity of diamond under extreme
  pressure: a first-principles study},
\newblock \bibinfo{journal}{Physical Review B} \bibinfo{volume}{86}
  (\bibinfo{year}{2012}) \bibinfo{pages}{115203}.
\bibitem[{Warren et~al.(1967)Warren, Yarnell, Dolling, and
  Cowley}]{warren1967lattice}
\bibinfo{author}{J.~Warren}, \bibinfo{author}{J.~Yarnell},
  \bibinfo{author}{G.~Dolling}, \bibinfo{author}{R.~Cowley},
\newblock \bibinfo{title}{Lattice dynamics of diamond},
\newblock \bibinfo{journal}{Physical Review} \bibinfo{volume}{158}
  (\bibinfo{year}{1967}) \bibinfo{pages}{805}.
\bibitem[{McSkimin et~al.(1972)McSkimin, Andreatch~Jr, and
  Glynn}]{mcskimin1972elastic}
\bibinfo{author}{H.~McSkimin}, \bibinfo{author}{P.~Andreatch~Jr},
  \bibinfo{author}{P.~Glynn},
\newblock \bibinfo{title}{The elastic stiffness moduli of diamond},
\newblock \bibinfo{journal}{Journal of Applied Physics} \bibinfo{volume}{43}
  (\bibinfo{year}{1972}) \bibinfo{pages}{985--987}.

\end{thebibliography}





\end{document}